\documentclass[preprintnumbers,aps,prd,floatfix,nofootinbib,twocolumn]{revtex4-1}

\pdfoutput=1

\usepackage{graphicx}
\usepackage{amsmath}
\usepackage{amssymb}
\usepackage{slashed}
\usepackage{braket}
\usepackage{mathtools}
\usepackage[utf8]{inputenc}
\usepackage{color}
\usepackage[usenames,dvipsnames]{xcolor}
\usepackage[normalem]{ulem}
\usepackage{soul}
\usepackage{units}
\usepackage{rotating}
\usepackage{hhline}
\usepackage{multirow}
\usepackage{tabularx}
\usepackage{bm}
\usepackage{longtable}
\usepackage{makecell}
\usepackage{setspace}

\usepackage{float}
\usepackage{multirow}
\usepackage[caption=false]{subfig}

\newcommand{\parz}[1]{\ensuremath{\left(#1\right)}}
\newcommand{\order}[1]{\ensuremath{O\parz{#1}}}

\newcommand{\eref}[1]{Eq.~(\ref{e.#1})}

\newcommand{\fref}[1]{Fig.~\ref{f.#1}}

\newcommand{\sref}[1]{Sec.~\ref{s.#1}}

\newcommand{\tref}[1]{Table~\ref{t.#1}}

\newcommand{\md}{M_{\text D}}
\newcommand{\gk}{g_{\text K}}
\newcommand{\qt}{q_{\text T}}
\newcommand{\pt}{P_{\text T}}
\newcommand{\bt}{b_{\text T}}
\newcommand{\btstar}{b_{\star}}
\newcommand{\btmax}{b_{\text{max}}}
\newcommand{\conv}{d \otimes {\cal C}  }
\newcommand{\zh}{z_h}
\newcommand{\chid}{\chi^2_{\text{d.o.f.}}}
\newcommand{\lorder}[1]{o\parz{#1}}
\newcommand{\gev}{\rm GeV}

\newcommand{\mk}{M_{\text K}}
\newcommand{\pk}{p_{\text K}}
\newcommand{\mo}{M_{\text 0}}
\newcommand{\mz}{M_{\text 1}}

\newcommand{\rows}[1]{\multirow{2}{*}{#1}}

\newcommand{\epm}{e^+e^-}

\allowdisplaybreaks
\begin{document}

\DeclareRobustCommand*\diff[2][]{%
   \mathop{
     \mathrm{d}^{#1}
     \mskip-0.2\thinmuskip
    #2}\nolimits
}
\newcommand{\di}[1]{\diff{#1}{}}

\newcommand{\intloop}{\int \frac{\diff{^{4 - 2 \epsilon} k}{}}{(2\pi)^{4 -2 \epsilon}}}
\newcommand{\intloopb}{\int \frac{\diff{^{4 - 2 \epsilon} l}{}}{(2\pi)^{4 -2 \epsilon}}}

\title{Transverse Momentum Dependent Fragmentation Functions from recent BELLE data.}

\author{M.~Boglione}
\email{mariaelena.boglione@unito.it }
\affiliation{Dipartimento di Fisica, Universit\`{a} di Torino, Via P. Giuria 1, 1-10125, Torino, Italy}
\affiliation{INFN - Sezione Torino, Via P. Giuria 1, 1-10125, Torino, Italy}
\author{J.~O.~Gonzalez-Hernandez}
\email{ joseosvaldo.gonzalezhernandez@unito.it}
\affiliation{Dipartimento di Fisica, Universit\`{a} di Torino, Via P. Giuria 1, 1-10125, Torino, Italy}
\affiliation{INFN - Sezione Torino, Via P. Giuria 1, 1-10125, Torino, Italy}
\author{A.~Simonelli}
\email{andrea.simonelli@unito.it}
\affiliation{Dipartimento di Fisica, Universit\`{a} di Torino, Via P. Giuria 1, 1-10125, Torino, Italy}
\affiliation{INFN - Sezione Torino, Via P. Giuria 1, 1-10125, Torino, Italy}
\date{2 July  2021}

\begin{abstract}
	
	A new formalism for the factorization of the cross section for single hadron production in $e^+e^-$ annihilations, differential in $z_h$, $P_T$ and thrust, is applied to the phenomenological analysis of data recently measured by the BELLE Collaboration. Within this scheme the $e^+e^- \to hX$ cross section can be recast in the convolution of a perturbatively calculable coefficient and a universal transverse momentum dependent fragmentation function. 
	While performing a next-to-leading order  calculation of the perturbative part of the process to next-to-leading logarithmic accuracy, we examine and thoroughly discuss the suitability of a number of possible ansatz to model the non-perturbative part of this universal transverse momentum dependent fragmentation function, showing the extent to which present experimental data can actually constrain its shape and functional form in terms of $z_h$, $P_T$ and thrust. 
\end{abstract}

\maketitle
\section{Introduction \label{sec:formalism}}

Transverse Momentum Dependent (TMD)  Parton Distributions (PDFs) and Fragmentation functions (FFs) are fundamental ingredients for the study of the inner structure of matter, as they  encode how fundamental constituents bind into hadrons and shed light on the hadronization mechanism that, thanks to the confinement properties of QCD, leads to the formation of hadronic states.
Their pivotal role in the investigation of the 3D structure of nucleons has motivated a huge effort in terms of experimental facilities as well as theoretical and phenomenological studies.

Unpolarized TMD PDFs are relatively well known objects, as their extraction can rely on combined analysis of different processes, like SIDIS and Drell-Yan scattering~\cite{Anselmino:2013vqa,Signori:2013mda,Bacchetta:2017gcc, Bacchetta:2019sam, Echevarria:2016scs},  for which dedicated TMD factorization theorems have been devised~\cite{Collins:2011zzd,Aybat:2011zv,Echevarria:2012js,Idilbi:2004vb}. 
On the contrary TMD FFs, their final state counterparts, are rather less known. 
In fact, the study of unpolarized TMD FFs is currently based on the phenomenological analysis of the sole SIDIS, as data for $\epm$ annihilations into two hadrons, the ideal framework for their determination, are not yet available.
To be precise, data on $\epm \to h_1 h_2 X$ processes are only available for polarized TMD FFs, like the pion and kaon Collins function, or for the $\lambda$ polarizing fragmentation function, for which several phenomenological studies have been performed; for example, some recent analyses can be found in  Refs.~\cite{Anselmino:2013vqa,Anselmino:2015sxa,Anselmino:2015fty,DAlesio:2020wjq}.
Moreover, extractions relying on SIDIS cross sections are inevitably affected by the strong 
correlation between the TMD PDF and the TMD FF which appear convoluted in the measured cross section. 
This issue could be circumvented by exploiting processes which involve only one TMD FF. In these regards, the thrust distribution of $\epm \to h\,X$, sensitive to the transverse momentum of the detected hadron with respect to the thrust axis, as recently measured by the BELLE collaboration~\cite{Seidl:2019jei}, is a very promising candidate, as it represents a process in which the TMD effects are traced back to one single hadron, observed in the final state. 
We note that some phenomenological analyses have been performed on $e^+e^-\to h X$ data~\cite{Boglione:2017jlh,Soleymaninia:2019jqo,Modarres:2021ffg}, where some subsets of  TASSO~\cite{Braunschweig:1990yd}, PLUTO~\cite{PLUTO:1983pce}, MARKII~\cite{Petersen:1987bq}, AMY~\cite{Bhattacharjee:1990iq}, CELLO~\cite{CELLO:1982fzq} data and the more recent BELLE~\cite{Seidl:2019jei} measurements have been considered.
These studies ignored or only partially addressed issues related to universality and factorization properties of $e^+e^-$ annihilations in a single hadron.
From a theory perspective, 
in fact, the study of the $\epm \to hX$ process has been very challenging, as 
standard TMD factorization techniques~\cite{Collins:2011zzd,Aybat:2011zv,Echevarria:2011epo,Echevarria:2012pw} do not apply.

As discussed in Refs.~\cite{Makris:2020ltr, Boglione:2021wov}, the $2$-jet final state topology of the above process can occur in three different kinematic configurations or ``Regions", denoted Region $1$, $2$ and $3$ in Refs.~\cite{Boglione:2021wov,Makris:2020ltr}, each corresponding to a different factorization theorem. 
These kinematic regions can be defined in terms of the size of the transverse momentum $P_T$ of the hadron observed inside the jet cone. 
If the hadron is detected very close to the thrust axis, the structure of the resulting factorization theorem is very similar to the standard TMD factorization, as in this case the soft radiation  significantly affects the transverse momentum of the detected hadron. This configuration corresponds to Region 1 and it has recently been investigated for pion~\cite{Kang:2020yqw} and $\Lambda$~\cite{Gamberg:2021iat,DAlesio:2020wjq} production 
neglecting the thrust dependence, which is integrated out.
On the other hand, if the hadron is detected very close to the jet boundary, its transverse momentum is large enough to affect directly the measured value of thrust. 
This configuration corresponds to Region 3. 
The associated factorization theorem involves a Generalized Fragmenting Jet Function (gFJF) rather than a TMD FF, and its treatment goes beyond the realm of TMD physics. 

While Regions $1$ and $3$ are rather extreme configurations of the $\epm \to h\,X$ phase space, the ``bulk" of events will belong to Region $2$,  
associated to the detection of hadrons with intermediate values of transverse momenta, neither extremely close to the thrust axis, nor too close to the jet external boundaries. Differently from the  two above kinematic configurations, the proper theoretical treatment of Region $2$ is still somehow controversial 
as the two main available approaches on this subject, Ref.~\cite{Makris:2020ltr} and Ref.~\cite{Boglione:2021wov}, do not find total  agreement on the final form of the corresponding factorization theorem. In this paper we will follow the factorization scheme devised in  Refs.~\cite{Boglione:2020cwn,Boglione:2020auc,Boglione:2021wov}, which offers some clear advantages for the practical implementation of a phenomenological analysis, leaving aside any discussion on the discrepancies between the two formalisms.  
These have been addressed in Section $5$ of  Ref.~\cite{Boglione:2021wov} and will be widely  discussed in a forthcoming paper \cite{Boglione-Simonelli:2022}. 

In Region $2$, soft radiation does not contribute actively to the generation of TMD effects. 
This is what makes the standard TMD factorization crucially different from 
the factorization mechanism of Region $2$, which shows features of both collinear and TMD factorization. The corresponding cross section can indeed be written as a convolution of a TMD FF with a ``partonic cross section", encoding the details of  thrust dependence. There are, however, two relevant issues that must be carefully taken into account. First of all, the TMD FF appearing in the $\epm \to hX$ factorized cross section of Region $2$ does not coincide with the usual TMD FF appearing in SIDIS cross sections. However, as we will discuss in more details below, differences between these two TMD definitions are well under control and their universality properties are not  undermined~\cite{Boglione:2020cwn}.
Hence, a phenomenological analysis of the thrust distribution of $\epm\to h\,X$ would allow to access the genuinely non-perturbative behavior of a TMD FF, 
free from any soft radiation effects. 

The second issue arises from the proper treatment of the rapidity divergences. Due to the very peculiar interplay between soft and collinear contributions, in Region 2  some of the rapidity divergences are naturally regulated by the thrust, $T$, but those associated with terms which are strictly TMD parts of the cross section need an extra artificial regulator, which is a rapidity cut-off in the Collins factorization formalism~\cite{Collins:2011zzd}. This induces a redundancy, which generates an additional relation between the regulator, the transverse momentum and thrust. Such relation inevitably spoils the picture in which the cross section factorizes into the convolution of a partonic cross section (encoding the whole $T$ dependence)  with a TMD FF (which encapsulates the whole $P_T$ dependence), as both these quantities turn out to depend on the rapidity cut-off. Hence, while the first becomes sensitive to the transverse momentum of the detected hadron, $P_T$, the other acquires a dependence on thrust, $T$. Moreover,  also the thrust resummation is intertwined with the transverse momentum dependence, making the treatment of the large $T$  behavior highly non-trivial.

A proper phenomenological analysis of Region $2$ must rely on a factorized cross section where the regularization of rapidity divergences is properly taken into account. As usual, all the difficulties encountered in the theoretical treatment get magnified in the phenomenological applications.
In this paper we will adopt some approximations, 
in order to simplify the structure of the factorization theorem without altering its main architecture. In particular, for single pion production from $\epm$ annihilation, we refer to the cross section presented in Ref.~\cite{Boglione:2020auc} 
\begin{widetext}
\begin{align} 
\label{eq:cross-sect}
& \frac{d \sigma}{dz_h \, dT \, d^2 \vec{P}_T} = 
-\sigma_B N_C 
\frac{\alpha_S}{4 \pi}  C_F 
\frac{3 + 8 \log{\tau}}{\tau} 
e^{
-\frac{\alpha_S}{4 \pi}\, 3  C_F 
\log^2{\tau} 
}
\sum_f \, e_f^2 \,
D_{1,\,\pi^{\pm}/f}(z_h,\,{P_T}/{z_h};\,Q,\,\tau\,Q^2) \,.
\end{align}
\end{widetext}
where $z_h$ is the fractional energy of the detected pion, $\tau = 1-T$ and $\sigma_B = {4\pi\alpha^2}/{3Q^2}$ is the Born cross section.

The unpolarized TMD FF, $D_{1,\,\pi^{\pm}/f}$, is defined in the impact parameter space, in terms of the transverse distance $\vec{b}_T$ Fourier conjugate of $\vec{q}_T \equiv {\vec{P}_T}/{z_h}$. At next-to-leading logarithmic (NLL) accuracy, and at the scales $\mu=Q$ and $\zeta=\tau Q^2$ as in Eq.~\eqref{eq:cross-sect}, it reads~\cite{Boglione:2021wov}:
\begin{widetext}
\begin{align}
&\widetilde{D}_{1,\,h/f}
(z_h,\,b_{\text{T}};\,Q,\,\tau\,Q^2) 
=
\notag \\
&\quad
\frac{1}{z_h^2}\bigg(
d_{h/f}(z_h,\,\mu_{b_\star}) + 
\frac{\alpha_S(\mu_{b_\star})}{4\pi}
\int_{z_h}^1 \, \frac{dz}{z}\,
\left[
d_{h/f}({z_h}/{z},\,\mu_{b_\star})
\,z^2 \, 
\mathcal{C}_{q / q}^{[1]}(z,b_*;\mu_{b_*},\mu_{b_*}^2) +
d_{h/g}({z_h}/{z},\,\mu_{b_\star})
\,z^2 \, 
\mathcal{C}_{g / q}^{[1]}(z,b_*;\mu_{b_*},\mu_{b_*}^2)
\right]
\bigg) 
\notag \\
&\quad \times
\mbox{exp} 
\left\{
\log{\frac{Q}{\mu_{b_\star}}}\,g_1 (\lambda) + g_2(\lambda)
+
\frac{1}{4} \, 
\log{\tau} \,
\left[
g_2^{K}(\lambda) + 
\frac{1}{\log{\frac{Q}{\mu_{b_\star}}}}\,g_3^{K}(\lambda)
\right]
\right\} 
\notag \\
&\quad \times
\md(z,b_{\text{T}}) 
\mbox{ exp} \left\{
-\frac{1}{4} \, \gk(b_{\text{T}}) \, 
\log{\left(\frac{Q^2}{M_H^2}\,\tau\right)}
\right \},
\label{e.tmd_NLL}
\end{align}
\end{widetext}
where the transition from small to large $b_{\text{T}}$ has been treated through  the $b_*$-prescription by defining 
\begin{align}
\label{eq:bstar}
b_\star \left(b_{\text{T}}\right) = \frac{b_{\text{T}}}{\sqrt{1 + (b_{\text{T}}/b_{\text{max}})^2}},\quad \mu_{b_*}=\frac{2e^{-\gamma_E}}{b_*}   
\end{align}
as is usual in the CSS formalism ~\cite{Collins:1984kg,Collins:1989gx,Collins:2011zzd}
.

Moreover, in order to ensure that integrating the above TMD FF renders the usual collinear FFs (indicated by lower-case $d$ in \eref{tmd_NLL}), we introduce in the $b_{\star}$-prescription a minimum value of $b_{\text{T}}$,
$b_{\text{min}}$, as in Ref.~\cite{Collins:2011zzd}, and replace Eq.~\eqref{eq:bstar} with $b_{\star} \left(\sqrt{b_{\text{T}}^2 + (b_{\text{min}}) ^2}\right)$. 
The first line of \eref{tmd_NLL} embeds the uppolarized TMD FF at short-distances and fixed scales $\mu = \mu_{b_\star} \equiv 2e^{-\gamma_E}/b_{\star}$ and $\zeta = \mu_{b_\star}^2$. It is a standard result to express this contribution as an operator product expansion where the operator basis are the collinear FFs and the Wilson coefficients are fully  predicted by perturbative QCD. 
The detailed expressions of the 1-loop Wilson coefficients are given in Appendix~\ref{app:tmd}.

The second line of \eref{tmd_NLL} describes the perturbative part of the evolution from $\mu = \mu_{b_\star}$ to $\mu = Q$ and from $\zeta = \mu_{b_\star}^2$ to $\zeta = \tau Q^2$. The functions $g_i$, $i=1,2$ and $g^K_j$, $j=2,3$ are required to reach the NLL-accuracy. They depend on the variable $\lambda = 2\,\beta_0 \, 
a_S(Q) \, \log{\frac{Q}{\mu_{b_\star}}}$. For convenience they are reported in Appendix~\ref{app:tmd}.

Finally, the last line of \eref{tmd_NLL} embeds the non-perturbative content of the unpolarized TMD FF, which is encoded in two non-perturbative functions, that must be extracted from experimental data. 
The first is the model function $\md$, which is the fingerprint of $D_{1,\,\pi^{\pm}/f}$ as it embeds the 
genuine large-distance behavior of the TMD. The second is the function $\gk$, describing the long-distance behavior of the Collins-Soper kernel, accounting for soft recoiling effects.
Notice that a factor $z_h$ is usually included~\cite{Collins:2011zzd} in the logarithm  of $\gk$, which is not present in \eref{tmd_NLL}. This simply corresponds to a different choice for the reference scale of evolution. We choose not to include it in order to have a $\gk$-factor completely unrelated to the $z_h$ dependence in $b_{\text{T}}$-space.
With respect to the usual definition of TMDs~\cite{Collins:2011zzd,Aybat:2011zv}, or ``square root definition" as labeled in Ref.~\cite{Boglione:2020cwn}, these two non-perturbative functions are related by the following equations
\begin{subequations}
\label{eq:tmddef_comparison}
\begin{align}
    \md^{\text{sqrt}}(z,b_{\text{T}}) &= \md(z,b_{\text{T}})\,\sqrt{M_{\text{S}}(b_{\text{T}})},
    \label{e.sqrtMD}\\
    \gk^{\text{sqrt}}(b_{\text{T}}) &= \frac{1}{2} \gk(b_{\text{T}}),
    \label{e.sqrtgK}
\end{align}
\end{subequations}
where $M_{\text{S}}$ is the soft model introduced in Ref.~\cite{Boglione:2020cwn}, describing the non-perturbative content of the soft factor appearing in standard TMD factorization theorems. Notice that while $\md$ is different in the two definitions, $\gk$ is basically the same, apart from a constant factor. Hence, for the extraction of $\gk$ from Region $2$ of $\epm \to h\,X$ we can test the parametrization already used in past phenomenological extractions, based on standard TMD factorization.
On the side of the TMD model, the comparison between the novel $\md$ extracted from Region $2$ of $\epm \to h\,X$ with its ``square root" counterpart 
will shed light on the soft model $M_{\text{S}}(b_{\text{T}})$, the remaining unknown required to perform global phenomenological analyses. 

The cross section in Eq.~\eqref{eq:cross-sect} can be obtained in two different ways. In Ref.~\cite{Boglione:2020auc} 
it is achieved by adopting a topology cut-off $\lambda$ that forces the cross section to describe a $2$-jet final state in the limit $\lambda \to 0$. This introduces an additional, artificial constraint which simplifies the computation of the transverse momentum dependent contributions by limiting the values of the transverse momentum to be smaller than the topology cut-off. Moreover, it allows to set an explicit relation linking the thrust, $T$, to the rapidity cut-off $\zeta$, namely  
$\zeta = \tau Q^2$. 
Finally, an approximated  
resummation of $\lambda$ produces the exponential suppressing factor of Eq.~\eqref{eq:cross-sect}, which replaces the effect of a proper thrust resummation~\cite{Boglione:2020auc}. Alternatively, Eq.~\eqref{eq:cross-sect} can be obtained from the correct factorization theorem of Region $2$  devised in Ref.~\cite{Boglione:2021wov} by making two rather strong  approximations. First, the whole transverse momentum dependence encoded \emph{outside} the TMD FF is integrated out up to the typical thrust-collinear scale $\sim \sqrt{\tau} Q$. This allows to recover the naive picture of a partonic cross section convoluted with a TMD FF. Then, the TMD is equipped with a rapidity cut-off, set to the minimal allowed rapidity for particles belonging to the same jet of the detected hadron, corresponding to $\zeta = \tau Q^2$. 
In this way, the underlying correlation between thrust and transverse momentum (due to the peculiar role of the rapidity regulator in Region $2$) is 
strongly simplified. 
Nevertheless, Eq.~\eqref{eq:cross-sect} embodies the 
essence of Region $2$, 
as the definition of the TMD FF is not affected by non-perturbative soft effects. Moreover, 
it represents the first attempt to account for the interplay between thrust and rapidity regulator.
In this paper, we present the first 
extraction of this universal TMD FF from $\epm \to h\,X$ data by the BELLE collaboration~\cite{Seidl:2019jei}, belonging to Region 2, within the specific framework of Refs. \cite{Boglione:2020auc,Boglione:2021wov}.

\section{Phenomenology}
In order to use Eq.~\eqref{eq:cross-sect}, complemented by the definition of unpolarized TMD FF in \eref{tmd_NLL}, 
one must choose 
parametric forms for $\md$ and $\gk$,  which describe the non-perturbative behavior of the TMD. Such choices are generally affected by the kinematical region of the data under consideration. 
This poses a big challenge since the error estimation of factorization theorems in QCD do not allow for sharp boundaries to be drawn. For instance, the small-$\qt$ cross section in  Eq.~\eqref{eq:cross-sect} and its associated  error of $\order{\qt^2/Q^2}$, do not  imply that the formalism should describe the data up to $\qt \sim Q$, but rather that in this region issues describing the data are to be expected. With no further indication of how far one can extend the description into the larger  $\qt$ region, one is left with model-dependent phenomenological results as the only indication of the validity of the formalism. An algorithm to delineate the contours of  $e^+e^- \to hX$ kinematic regions where specific factorization regimes can be applied was developed in Ref.~\cite{Boglione:2021wov}, which we will refer to in our analysis.
Another delicate point is the choice of collinear fragmentation functions. While one expects  part of the $z$-dependence of theory lines to come from the behavior of the collinear FFs, there is no restriction regarding a possible $z$-dependence in the function $\md$. Again, how appropriate a given set is depends on the parametric form of the model. In the following sections we systematically  explain our choices.

For our study we will use  a simple minimization procedure of the $\chi^2$  given
by
\begin{align}
\label{e.chi2}
\chi^2=\sum_{j=1}^{n}\frac{(T_j(\{p\})-E_j)^2}{\sigma_j^2}\,,
\end{align}
with $\{E_j\}$ the set of the $n$ data points under consideration and 
where the corresponding theory computations $\{T_j\}$ depend on a set $\{p\}$ of  $m$ parameters. The uncertainties $\sigma_j$ are treated as independent uncorrelated errors, i.e. different sources of errors provided by the BELLE data set are added in quadrature. Future refinements of our work can be achieved by modifying the definition in \eref{chi2} in order to account for the correlations in the systematic uncertainties. This, however, requires more detailed information about the different sources of such types of errors, which is not available. For now, we proceed by minimizing \eref{chi2} as done in previous related analyses \cite{Bertone:2017tyb,Soleymaninia:2019jqo,Kang:2020yqw,DAlesio:2021dcx,Gamberg:2021iat}.

In order to test goodness-of-fit, we use the  $\chi^2$ per degree of freedom, given by  $\chid=\chi^2/(n-m)$, which should be close to unity for a model to be considered appropriate. We will estimate the statistical errors of our analysis by determining  $2\sigma$-confidence regions based on a straight forward application of the Neyman-Pearson Lemma and Wilks' theorem. Concretely, provided a minimal set of parameters $\{p_0\}$ with $\chi^2_0$, we consider  parameter configurations $\{p_i\}$ with $\chi^2_i$ given by
\begin{align}
\chi^2_i<\chi^2_0+\Delta\chi^2\,,
\end{align}
where $\Delta\chi^2$ is \emph{not} an arbitrary tolerance but rather depends on the confidence level and the number of parameters varied. For $c$-$\sigma$ confidence level one has
\begin{align}
\text{erf}\left(\frac{c}{\sqrt{2}}\right)=\int_{0}^{\Delta\chi^2} dx\,\, X^2_{D}(x)\,,
\end{align}
with $X^2(D)$ a chi squared  distribution with $D$ degrees of freedom equal to the number of parameters varied.\footnote{This equation gives $\Delta\chi^2=1$ for $1\sigma$ c.l. when varying only one parameter. We consider $2\sigma$ and mostly vary all parameters at once so $\Delta\chi^2$ values will be larger than unity.}
\subsection{TMD FF z-dependence and choice of collinear FFs}
\label{s.zdep}
 Similarly to the usual CSS formalism for two-hadron production, the  impact parameter space in \eref{tmd_NLL} is constrained at small $\bt$  by a small distance OPE, hence the appearance of the convolution of collinear FFs with matching coefficients $\cal C$, which we denote by $\conv$. This factor provides an important constraint of the $\zh$-dependence for the TMDs. As discussed before, the transition from short to large distance of the TMD is regulated by the $\btstar$-prescription, for which a maximum value or ``freezing point" must be set, below  which one expects perturbation theory to apply. Such maximum distance, $\btmax$ in Eq.~\eqref{eq:bstar}, corresponds to a minimum perturbative scale of $\mu_{min}=2e^{-\gamma_E}/\btmax$.
 For our studies we choose $\btmax=1.0\, \text{GeV}^{-1}$, which ensures that perturbative quantities are never evaluated bellow a scale of $1.12\,\text{GeV}$. This seems like a sensible choice since  perturbation theory is known to work well in collinear observables down to a scale of around $1.0\, \text{GeV}$.

 With this choice, we turn to the question of choosing a set of collinear FFs. We will compare the NNFF\cite{Bertone:2017tyb}  and the JAM20\cite{Moffat:2021dji} next-to-leading order (NLO) sets 
 \footnote{Note that we use a recent update of the JAM20 pion FFs, obtained from https://github.com/QCDHUB/JAM20SIDIS.}. 
 These are modern analyses that represent the state of the  art in collinear FF extractions and are readily available through  LHAPDF~\cite{Buckley:2014ana}. As it can be seen in \fref{conv1}, computation  of $\conv$ may render significantly different results for each collinear FF set. One may suspect that the extraction of the TMD is sensitive to the choice of collinear functions. It is however not obvious that either of the collinear set is to be preferred over the other. It is entirely possible that by adjusting values of the model  parameters for say, $\md$, a similar description of the data could be achieved with the two collinear FF sets. By any consideration, the question of which set is more appropriate  depends on the choices of the model. 
 
 \begin{figure}
 	\centering
 	\includegraphics[scale=1]{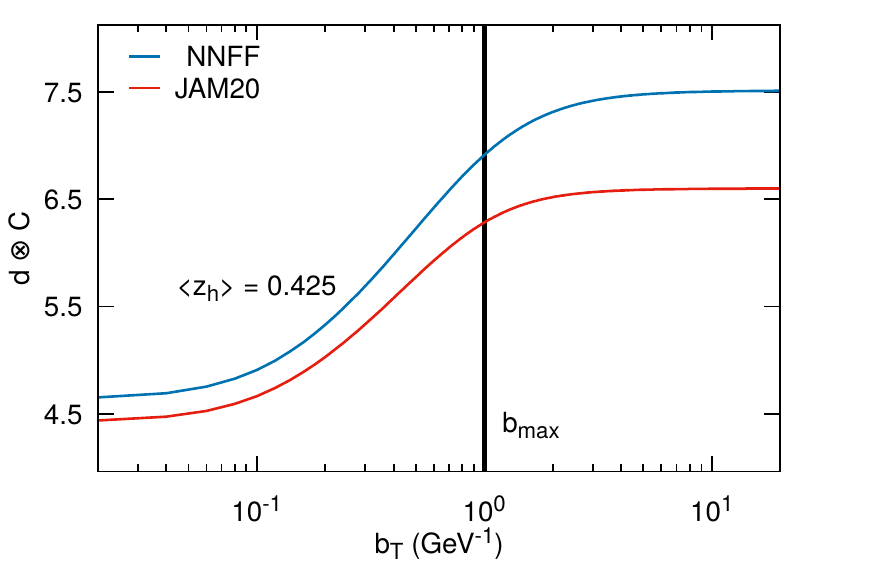}
 	\caption{
 		\label{f.conv1}
 		Convolution of the collinear fragmentation function and matching coefficients $\conv$ for the NNFF\cite{Bertone:2017tyb} and JAM20\cite{Moffat:2021dji} sets. Here $z$ is fixed at $z=0.425$, but significant differences can also be observed at other values of $\zh$.}
 \end{figure}
 
In order to choose a set, we  perform preliminary fits at fixed values of $T=0.875$ and look for the one that better describes the data, in terms of the minimal $\chi^2_{\text {dof}}$. We consider for now only the kinematical ranges $0.375~<~\zh~<~0.725$ and $\qt/Q<0.20$. This includes enough data points to constrain the tests. At this stage we only attempt to  parametrize  $\md$ and set the exponential factor containing $\gk$, in last line of \eref{tmd_NLL}, equal to unity.

Notice that according to Ref.~\cite{Boglione:2021wov} data corresponding to  $z_h$ bins with $z \le 0.375$ would be dominated by Region 1, which requires a different factorization theorem. For this reason we do not consider them here.

\begin{table}
	\caption{Models in impact parameter space used for preliminary tests in this section. First two entries correspond to $\zh$-independent models for $\md$. Models labeled as "BK" are proportional to a modified Bessel function of the second kind and correspond to a power law in momentum space. Entries three and four  are $\zh$-dependent models for $\md$, obtained by modifying the mass parameter of the BK model, as indicated. The last entry introduces $\zh$-dependence to the BK model by a multiplicative factor with Gaussian behavior in $\bt$.}
	\label{t.models1}
	\begin{center}
		\begin{tabular}{|l|c|c|}
			\hline
			ID&$\md$-model&parameters\\
			\hline
			\hline
			\multicolumn{3}{|c|}{$\zh$-independent models}\\\hline
			\multirow{3}{*}{1)Exp-p}&\multirow{3}{*}{$e^{-(\mo \bt)^p}$}&\multirow{3}{*}{$\mo,\,p$}\\ 
			& & \\
			& & \\
			\hline
			\multirow{3}{*}{2)BK}&\multirow{3}{*}{$\cfrac{2^{2-p} (\bt \mo)^{p-1} }{\Gamma (p-1)}K_{p-1}(\bt \mo)$}&\multirow{3}{*}{$\mo,\,p$}\\ 
			& & \\
			& & \\
			\hline
			\multicolumn{3}{|c|}{$\zh$-dependent models}\\\hline
			\multirow{3}{*}{3)BK-1}&\multirow{3}{*}{$ \mo\to M_1\left(1-\eta_1 \log (\zh)\right)$}&\multirow{3}{*}{$M_1,\,\eta_1,\,p$}\\ 
			& & \\
			& & \\
			\hline
			\multirow{3}{*}{4)BK-2}&\multirow{3}{*}{$ \mo\to M_2\left(1+\cfrac{\eta_2}{\zh^2}\right)$}&\multirow{3}{*}{$M_2,\,\eta_2,\,p$}\\ 
			& & \\
			& & \\
			\hline
			\multirow{3}{*}{5)BK-g}&\multirow{3}{*}{$e^{(M_g\bt)^2 \log(\zh)} \times $ BK}&\multirow{3}{*}{$M_g,\,\mo,\,p$}\\ 
			& & \\
			& & \\
			\hline
		\end{tabular}
	\end{center}
\end{table}
In a first attempt to test the  collinear functions, one may consider models for $\md$ with no explicit $\zh$-dependence, and perform fits for fixed values of $\zh$.  The choice of models is summarized in the top two entries of \tref{models1}:  
model $1$ inspired by a Gaussian-like $b_{\text{T}}$ behavior while model 2, proportional to a modified Bessel function of the second kind, corresponds to a power law in momentum space and is the same functional form considered for  $\md$ in  \cite{Boglione:2020auc}.  
As it can be seen in \fref{chi21},  these models result in rather high values of $\chi^2_{\text{dof}}$, giving a bad description of the data. Nonetheless, it is noteworthy that the $\chi^2_{\text{dof}}$ tends to be larger for the JAM20 set. Both models seem to work at $\qt/Q<0.1$ but deteriorate fast for $0.1<\qt/Q<0.2$. In the following sub-sections we  will set our final $\qt$-cut to the intermediate value $\qt/Q<0.15$. For now we will leave this aside and continue to address the $\zh$-dependence.
\begin{figure}
	\centering
		\includegraphics[scale=0.9]{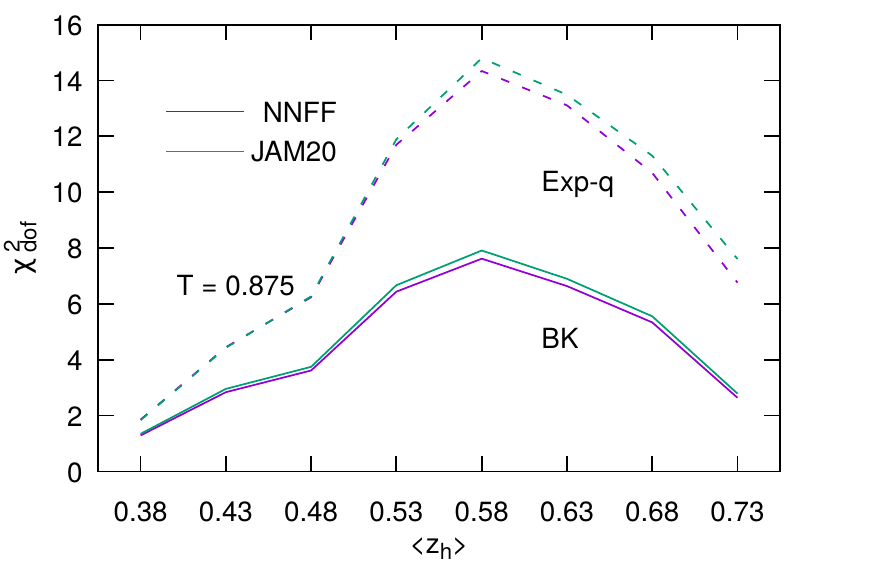}
	\caption{
		\label{f.chi21}
		 Minimal $\chi^2_{\text{dof}}$ for fits at fixed $T=0.875$ and individual $\zh$-bins  in the range $0.375~<~\zh~<~0.725$, for $\md$ models with no $\zh$-dependence. Here $q_T/Q<0.20$. Dashed and solid lines correspond respectively to the first and second entries in \tref{models1}. For each model we have two parameters and a total of nine individual  fits, one per $\zh$-bin. 
		 Note that even with such large values of $\chi^2_{\text{dof}}$, the mild relative differences between the using JAM20 and NNFF suggest that either set could describe the data to the same quality.} 
\end{figure}
 Recall that so far we have performed only independent fits at fixed $T=0.875$ and separately for each bin inside the range $0.375~<~\zh~<~0.725$. A useful exercise is to  plot the values of the resulting minimal  parameters in terms of $\zh$, as is done in \fref{para1}, for the BK model. There, it is clear that if one expects to fit all bins in $z_h$ simultaneously (still at fixed $T=0.875$), some $\zh$-dependence shall be needed in the parametric form for $\md$. We remark that an important  result of the factorization scheme is that $\gk$ must be independent of $\zh$. 
 Another interesting aspect of \fref{para1} is that a stronger $\zh$-dependence is observed for the mass parameter $M$ than for the dimensionless parameter $p$. We find that improving  the trend of theory lines in the variable $\zh$ is more readily done by introducing a  $\zh$-dependence in  dimensionful  parameters. We have observed this for several cases we tested, although here we only show a few of them. More generally, one could expect  strong correlation between all parameters in $\md(b_{\text{T}})$. For instance, a closer inspection of the example in \fref{para1} shows that the two parameters shaping the $b_{\text{T}}$ profile of $\md$, $M_0$ and $p$, display a similar trend as a function of $\zh$. We will come back on this later on in the next sections.

 \begin{figure}[t]
 	\centering
 	\includegraphics[scale=0.9]{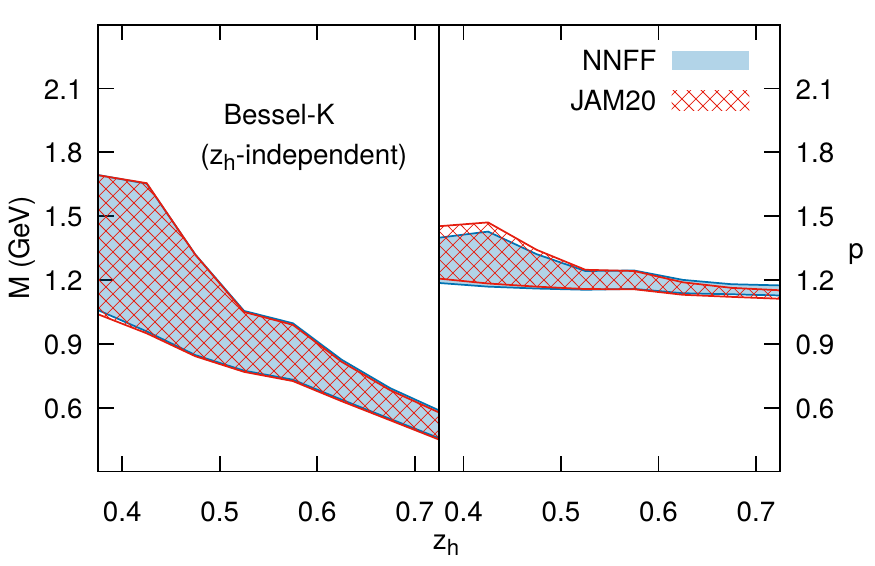}
 	\caption{
 		\label{f.para1}
 		Minimal parameter values for fits at fixed $T=0.875$ and individual $\zh$-bins  in the range $0.375~<~\zh~<~0.725$, for the $\md$ model in the second entry of \tref{models1} ($\zh$-independent BK model). Here $q_T/Q=0.15$. Results correspond to the solid lines in \fref{chi21}. In this case, where we fit $\zh$-bins separately, the incompatibility of $M$ and $p$ for different $\zh$ suggests that a $\zh$-dependence is needed if the model is to describe the data on a simultaneous fit of the $0.375~<~\zh~<~0.725$ range. It is interesting to note that the dimensionful parameter $M$ exhibits a stronger correlation to $\zh$.
 		}
 \end{figure}

 We attempt three different $\zh$-dependent models for $\md$, as indicated in the last  three entries of \tref{models1}. The first two are  modifications of the BK model, where we modify the mass parameter as $M\to M(z)$, adding in each case one more parameter to introduce, respectively, a linear and a logarithmic term. The last one, is  the BK model multiplied  by  $\zh^{(M_g \bt)^2}$, so that the $\zh$-dependence is controlled by this additional multiplicative function and determined by the mass parameter $M_g$. \\
 Results for these three models can be seen on the left panel of \fref{chi22}. 
 Despite the large values of $\chi^2_{\text{dof}}$ for the first two models, we find a considerable improvement with respect to the $\zh$-independent BK model. The third model works indeed much better, which is partly due to its  $\zh$-dependence but also to the Gaussian behavior introduced by the factor $\zh^{(M_g \bt)^2}$. The Gaussian behavior of this model improves the description  at the large end of the selected range of $\qt$, giving much lower values of $\chi^2_{\text{dof}}$. For this last model, last entry in \tref{models1}, we perform two more fixed-$T$ fits  for $T=0.750$ and $T=0.825$, resulting in $\chi^2$'s roughly three times smaller than those corresponding to models BK1 and BK2. Results are shown on the right panel of \fref{chi22}.
 
 One should be careful to interpret these results. First, while it may seem that the last model should be the obvious choice to extract the unpolarized , the other two  $\zh$-dependent models we have considered here are able to describe the data well up to $\qt/Q<0.1$, as we will show in the following sub-sections. This is a delicate point, since one does not know a priori for which maximum value of $\qt/Q$ one can still trust that the errors $\order{(\qt/Q)^2}$ of \eref{tmd_NLL} are small  enough so that the formalism is still valid. For instance, if the  cut on  $\qt/Q$ was made more restrictive, say $\qt/Q < 0.1$,  
 the clear advantage of the Gaussian $\zh$-dependent model, describing the data in the region  $0.1<\qt/Q<0.2$, would become less significant.

 We close our preliminary  discussion of the $\zh$-dependence by stating the main conclusions of this subsection. 
 First, a stronger $\zh$-dependence is observed in mass parameters than in dimensionless parameters. 
 This is an observation that applies to several models we tested, of which we provide one concrete  example in \fref{para1}. 
 In the specific case of \fref{para1}, we also find that $\zh$ may strongly correlate the model parameters $M_0$ and $p$. 
 Second, in all the preceding  discussions, and despite of  inadequacies in  some of the models considered, $\chi^2_{\text{dof}}$ values tend to be smaller with NNFF, 
 so this will be our choice for our main analysis, but we will not yet set on a specific  model for $\md$. Based on our preliminary studies of this section, we expect that using  JAM20 would give larger values of $\chi^2_{\text{dof}}$, although not by much.

 \begin{figure}
 	\centering
 	\includegraphics[scale=0.9]{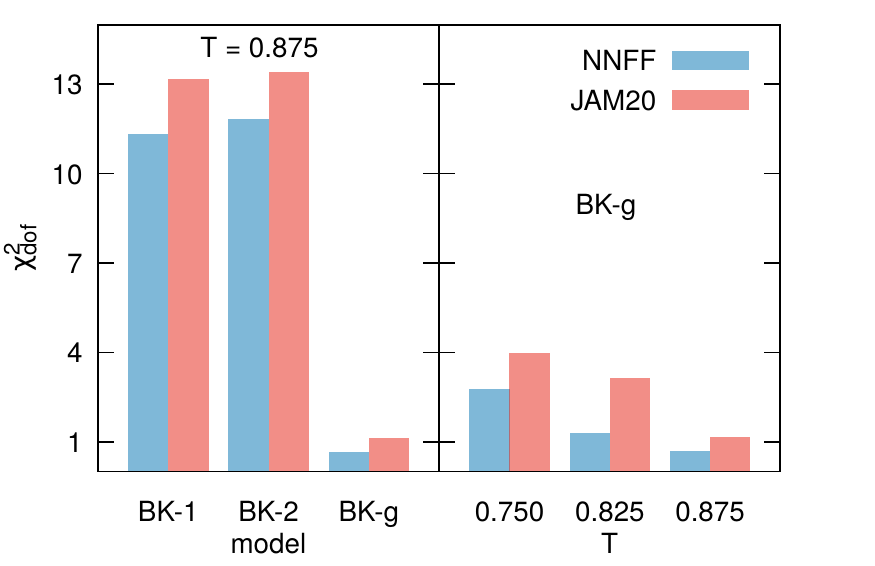}
 	\caption{
 		\label{f.chi22}
 		Minimal $\chi^2_{\text{dof}}$ for fits in the kinematic range $0.375~<~\zh~<~0.725$ ($\zh$-bins are fitted simultaneously), for the $\zh$-dependent models for $\md$ in the last three entries of \tref{models1}. Left panel: comparison of the results obtained with NNFF\cite{Bertone:2017tyb} and JAM20\cite{Moffat:2021dji}, for fixed $T=0.875$. Right panel: fixed-$T$ fits for $T=\{0.750,0.825,0.875\}$, using the BK model with a gaussian $\zh$-dependent term (last entry in \tref{models1}). Similarly to the results presented in \fref{chi21}, the NNFF consistently produce smaller  values of $\chi^2_{\text{dof}}$.
 				}
\end{figure}

\subsection{ Behavior of the unpolarized TMD FF in the large-$\bt$ limit. }\label{s.largebT}

In this subsection we will address the behavior of the unpolarized TMD FF in impact parameter space. Specifically, we look at possible parametric forms for $\md$ in \eref{tmd_NLL},  paying special attention to the large-$\bt$ limit. For the purposes of our discussion we identify two different possible meanings for "large-$\bt$" behavior:
\begin{enumerate}
\item asymptotically large-$\bt$
\item maximum $\bt$ accessible through data. 
\end{enumerate}  
The first one corresponds to the formal limit $\bt\to\infty$, in which one may write  asymptotic expansions for a known parametric form. For instance, the BK model discussed in the previous subsection has an asymptotic limit
\begin{align}
	\frac{2^{2-p} (\bt \mo)^{p-1} }{\Gamma (p-1)}&K_{p-1}(\bt \mo)\nonumber\\
	&\to
	\sqrt{\pi }\frac{ 2^{\frac{3}{2}-p}  (\bt M)^{p-\frac{3}{2}}}{\Gamma (p-1)}e^{-\bt \mo}\,.
	\label{e.BK}
\end{align}
characterized by an exponentially decaying behavior as $\bt\to\infty$. The second one, instead, refers  to the largest region in $\bt$ that is accessible phenomenologically, i.e., the largest distances at  which the data can constrain the model, which can be better determined after carrying out a data analysis. 
 The largest $b_{\text{T}}$ accessible phenomenologically  corresponds to the case of measurements at values of $Q$ small enough that nonperturbative effects are maximized, but large enough that TMD factorization still holds. Even at scales of, say, $Q=2\,\text{GeV}$, it is possible that the asymptotic behaviour of the TMDs cannot be resolved completely. At BELLE kinematics, where $Q\approx10\,\text{GeV}$, it is unlikely that one can find strong constraints for the asymptotic behaviour of TMDs. 
 
 \begin{table}
	\caption{Models for $\md$ in impact parameter space. Both cases shown are obtained by multiplying model BK of \tref{models1}, which corresponds to a power law in momentum space, by an additional function of $\bt$ and $\zh$.  }
	\label{t.models2}
	\begin{center}
		\begin{tabular}{|c|l|c|}
			\hline
			\multicolumn{3}{|c|}{\multirow{2}{*}{}}\\
			\multicolumn{3}{|l|}{\multirow{2}{*}{$\quad\md=\cfrac{2^{2-p} (\bt \mo)^{p-1} }{\Gamma (p-1)}K_{p-1}(\bt \mo)\,\,\times\,\,F(\bt,\zh)$}}\\
			\multicolumn{3}{|c|}{\multirow{2}{*}{}}\\
			\multicolumn{3}{|c|}{\multirow{2}{*}{}}\\
			\multicolumn{3}{|l|}{\multirow{2}{*}{$\quad M_z=-\mz\log(\zh)$}}\\
			\multicolumn{3}{|c|}{\multirow{2}{*}{}}\\
			\multicolumn{3}{|c|}{\multirow{2}{*}{}}\\
			\hline
			ID&$\qquad\qquad F$-model&parameters\\
			\hline
			\hline
			\multirow{3}{*}{ I }&\multirow{3}{*}{$F=\left(\cfrac{1+\log\left(1+(\bt M_z)^2\right)}{1+\left((\bt M_z)^2\right)}\right)^q$}&\multirow{3}{*}{$\mo,\mz,p,\,q=8$}\\
			& & \\
			& & \\
			\hline
			\multirow{3}{*}{ IG }&\multirow{3}{*}{$F=\exp\left((M_g\bt)^2 \log(\zh)\right)$}&\multirow{3}{*}{$\mo,\,M_g,\,p$}\\ 
			& & \\
			& & \\
			\hline
		\end{tabular}
	\end{center}
\end{table}

This would mean that fitting BELLE data may be possible with parametric forms of distinct asymptotic behaviour. 
However, when considering data at smaller energy scales, for which the maximum $\bt$ accessible is likely larger than that at BELLE energies, one may find inconsistencies in a global fit if the asymptotic behaviour of $\bt$ is not chosen appropriately Theoretical constraints are important in light of all these issues encountered at lower energy phenomenology, see for example Refs.~\cite{Boglione:2014oea,Collins:2016hqq,Gonzalez-Hernandez:2018ipj,Wang:2019bvb,Boglione:2016bph,Boglione:2019nwk,Boglione:2022gpv}. To do so, we follow some of the considerations made in Ref.~\cite{Collins:2014jpa}. Thus, for this work we will look for a parametric $\md$ that in $\bt$ space decays exponentially, but that is able to describe BELLE data at least as well as  model 5 in \tref{models1}, which in the preliminary cases considered so far, seems to be suitable.  A possible candidate is shown in  \tref{models2}, where for convenience we have explicitly rewritten model 5 of \tref{models1}. Both models in \tref{models2} correspond to a power-like behaviour in momentum space, characterized in $\bt$ space by the modified Bessel function of the second kind, times an extra factor which we denote as $F$. To make the comparison between exponential and Gaussian asymptotic behaviour more transparent, in this preliminary study we consider only the models in \tref{models2}. Note that even in the case  $F=1$ one may recover an exponentially decaying behaviour asymptotically, from the Bessel function alone, as seen in \eref{BK}. We will consider this case  later as it requires a detailed explanation of possible final parametric forms, which account for the strong correlations of parameters in $\md$ related to the $\zh$ dependence, as noted in the previous subsection.

For now, we will compare how well the models in \tref{models2} may describe the data. 
Our aim is to provide a practical example where two models that describe the data reasonably well, are not necessarily constrained in the asymptotically large $\bt$ limit. Decoupling the question of what is an appropriate parametric form for the $\pt$  behaviour of $\md$ is not independent of the choices to model its $\zh$ dependence. Thus we proceed as follows.
First, we perform three fits at fixed values of $T\;=\;\{0.750,0.825,0.875\}$, 
where in each case, we include BELLE data in the region  $\qt/Q<0.20$ and $0.375<\zh<0.725$. To accommodate the $\zh$ dependence we choose a logarithmic behaviour in the function $F$ as shown in \tref{models2}. Since we are not fitting the three $T$ bins simultaneously, we will not be able to also fit $\gk$, which correlates to thrust, so for now we set $\gk=0$. Then, we will look at a single case, one value of $T$ and $\zh$, where the $\pt$ dependence is described well by both models and look at the results in  $\pt$ and $\bt$ space. 

\begin{table}
	\caption{
	Minimal $\chid$ resulting by fitting the two para\-metric forms for $\md$ in \tref{models2}. In each case we perform three independent fits, one for each value  $T=\{0.750,0.825,0.875\}$, in the ranges $\qt/Q<0.2$ and $0.375<\zh<0.725$. As far as the description of the data is con\-cerned all three cases seem to be acceptable, see explanation in the text.  }
	\label{t.chi21}
	\begin{center}
		\begin{tabular}{c c c c c}
			\hline
			\multicolumn{5}{c}{$\chid$ (fixed-$T$ fits)}\\
			\hline
			&&&&\\
			$\md$ model  &$\,\,\,T=$&0.750&0.825&0.875\\
			\hline
			&&&&\\
			\hline
			&&&&\\
			I &	&1.2&0.38&1.02\\
			&&&&\\
			IG &	&1.46&0.47&1.51\\
			&&&&\\
			\hline
		\end{tabular}
	\end{center} §
\end{table}

The results of the fixed-$T$ fits are shown in \tref{chi21}. The smaller values of $\chid$ obtained with model I are related to the choice $q=8$, which allows for a good description of the $\zh$ bins considered.  Note that modifying the $\zh$ behaviour in model IG could improve its best fit $\chid$ as well. At this stage we consider both models as candidates to parametrize $\md$, since our main interest is to discuss about the $\pt$ dependence. 
\begin{figure}
	\centering
	\includegraphics[scale=0.9]{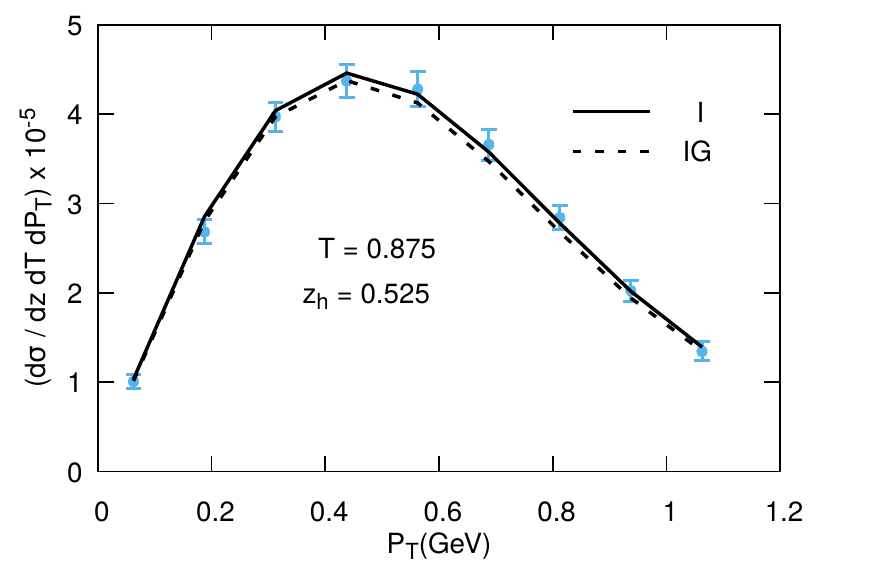}
	\caption{ 
		\label{f.mdvsqt1}
		Best-fit lines for both models in \tref{models2}, obtained by fitting BELLE data for the kinematics $T=0.825$, $0.375<\zh<0.725$ and $\pt/ \zh Q<0.2$. Note that both lines follow essentially the same profile in the region of the data shown.
	}
\end{figure}

Now we look at the case $\zh=0.525$ and $T=0.825$, for which both models describe the data reasonably well. In fact, as seen in \fref{mdvsqt1}, the models of \tref{models2} have the same profile and almost lie on top of each other. Corresponding lines in $\bt$ space are shown in \fref{mdvsbt1}, where it can be seen that for values $\bt>4$ GeV$^{-1}$ the cross section calculated using models I and IG deviate.  This is of course due to the differences in the asymptotic behaviour of the models. This example simply illustrates that the asymptotic behaviour of the TMD ff is not necessarily constrained by BELLE data after some large value of $\bt$. However, the  reason to prefer an asymptotic behaviour like that of model I comes from the necessity to fit data at lower energies in the future, for which the large-$\bt$ Gaussian fall off may not be appropriate. 

From here on out we will focus on models for $\md$ that decay  exponentially  in the asymptotically large $\bt$ limit. More precisely
\begin{align}
\label{e.modelasyconst}
	\qquad\log(\md)\underset{\bt\to\infty}{\sim} - \,C \,\bt + \lorder\bt,\,
\end{align}
with $C$ a positive mass parameter and where we have used the little-$o$ symbol to indicate sub-linear terms in $\bt$. 
Furthermore, we will explore two different approaches, leading to two classes of models. 
The first one is model I in table \tref{models2}, which corresponds to the function of \eref{BK} times the $\zh$-dependent function $F$. The second one, is similar to model I but sets $F=1$ and models the $\zh$ dependence through both the mass parameter $\mo$ and the power $p$ of the Bessel function function of \eref{BK}.

\begin{figure}[t]
	\centering
	\includegraphics[scale=0.9]{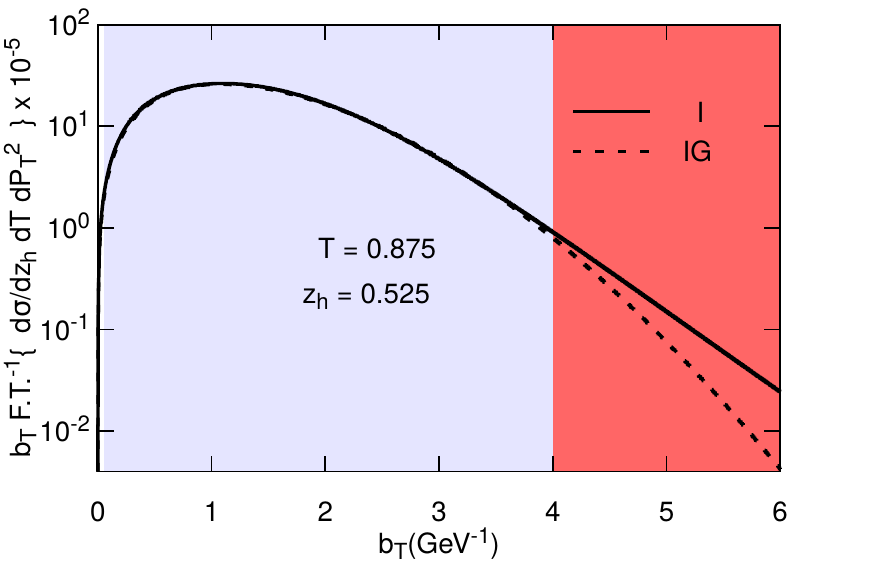}
	\caption{ 
		\label{f.mdvsbt1}
		Best-fit lines for both models in \tref{models2} in $\bt$ space, obtained by fitting BELLE data for the kinematics $T=0.825$, $0.375<\zh<0.725$ and $\qt/\zh Q<0.2$. Lines correspond to those in \fref{mdvsqt1}. The deviation of the two theory lines after $\bt>4\,\text{GeV}^{-1}$ indicates the lack of sensitivity to the asymptotic behaviour of the models in this particular example.
	}
\end{figure}

Before performing our extraction, however, we need to 
set a parametric form for $\gk$.

\subsection{Behavior of $\gk$ in the large-$\bt$ limit. }

The usual definition of the TMD FF in the CSS formalism differs from that introduced in Ref.~\cite{Boglione:2020cwn} by a non-perturbative function $M_{\text{S}}(b_{\text{T}})$, as explained in Section~\ref{sec:formalism} and given in \eref{sqrtMD}. $M_{\text{S}}(b_{\text{T}})$  is associated to soft gluon effects and originates from the fact that in the latter definition the TMDs are purely collinear objects, while in the CSS definition soft radiation contributions are included in the TMD definition itself. 
This means that the non-perturvative function $\md(b_{\text{T}})$ introduced in Eq. \eqref{eq:cross-sect}, and discussed in Section \ref{sec:formalism}, cannot be used directly in $e^+e^-$-two hadron production  or  SIDIS processes, see \eref{sqrtMD}. Note, however, that the non-perturbative function $\gk$  has been defined to be the same as in the usual CSS formalism, up to a trivial factor of $2$, see \eref{sqrtgK}. Thus, it characterizes the large distance behavior of the Collins-Soper kernel as defined in \cite{Collins:2011zzd}. This is perhaps one of the most useful aspect of the formalism in Refs.~\cite{Boglione:2020auc,Boglione:2020cwn,Boglione:2021vug,Boglione:2021wov}  in the context of global fits,  since it allows for comparisons of the extracted $\gk$ with  other recent  work (see for example Refs. \cite{Bacchetta:2017gcc,Bacchetta:2019sam,Scimemi:2017etj,Scimemi:2019cmh}).  In order to choose a suitable parametrization for $\gk$, we use the following observation as a guiding principle.

In general, one may write the TMD FF in $\bt$  space as 
\begin{align}
	\tilde{D}(\bt,\zeta)=&\tilde{D}(\bt,\zeta_0) \exp\left\{-\frac{\gk}{4} \log\left(\frac{\zeta}{\zeta_0}\right)\right\}\big(...\big)\,,
\end{align}
where only the dependence on $\bt$ and $\zeta$ has been written explicitly, and the ellipsis indicate other terms containing perturbatively calculable quantities.  Using the hypothesis in \eref{modelasyconst} one has that  in the large-$\bt$ limit
\begin{align}
\log\left(\tilde{D}(\bt,\zeta)\right)\overset{\bt\to\infty}{=}&-C \bt -\frac{\gk^{\text{large} \, \bt}}{4} \log\left(\frac{\zeta}{\zeta_0}\right)+ o(\bt)\,.
\end{align}
We then note that
\begin{align}
\label{e.gkasy}
\gk\overset{\bt\to\infty}{=}o(\bt)\implies\log(\tilde{D}(\bt,\zeta))=O(\bt)\,,
\end{align}
 independently of $\zeta$ and $\zeta_0$. This seems like  a reasonable condition since $\zeta_0$ is a somewhat arbitrary reference scale: for instance, it could be chosen depending on the kinematics of a particular phenomenological analysis. We will consider in this analysis only the hypothesis that asymptotically $\gk=o(\bt)$. As a counter example, with the same ansatz for the asymptotic behavior of  $\tilde{D}(\bt,\zeta)$, \eref{modelasyconst}, choosing the large-$\bt$ behavior of $\gk$ to be quadratic would implicitly assign a special role to the reference scale $\zeta_0$, in the sense that in this case  for  $\zeta=\zeta_0$,  $\log(\tilde{D}(\bt,\zeta))=O(\bt)$, while for $\zeta\neq\zeta_0$, $\log(\tilde{D}(\bt,\zeta))=O(\bt^2)$. Note that one could 
 set $\gk$ to be $O(\bt)$ instead of $o(\bt)$ and still have \eref{gkasy} be valid. However, this allows for $\tilde{D}(\bt,\zeta)$ to be divergent  in the limit $\bt\to\infty$, for sufficiently small $\zeta/\zeta_0$ (see also the discussion in Ref.~\cite{Collins:2014jpa}). Note that a sub-linear $\bt$ behaviour for $\gk$ has already been suggested by several authors, see for instance  Eq.~(79) in \cite{Collins:2014jpa}, Eq.~(40) in \cite{Aidala:2014hva} and Eq.~(24) in \cite{Vladimirov:2020umg}).

Our analysis will be conducted by adopting the following functional forms for the large $\bt$ behaviour of $\gk$
\begin{align}
	&
	\gk\overset{\bt\to\infty}{\sim}
	\log(\mk\bt)
	\label{e.gk-largeb-jogh}\\
	&
 	gk\overset{\bt\to\infty}{\sim}
	(\mk \bt)^{(1-2\pk)},\qquad 0<\pk<1/2\,\label{e.gk-largeb-vlad}
\end{align}
where the first expression is similar to that considered in Ref. \cite{Aidala:2014hva} (but with an undetermined power $p_k$), while the second expression corresponds to the model calculation presented in  Ref.~\cite{Vladimirov:2020umg} for the CS kernel as $\bt\to\infty$. We have also considered a constant asymptotic form, as suggested in Ref.~\cite{Collins:2016hqq} but, limited to the data sample we are presently fitting, we obtain  consistently larger $\chi^2$s compared to those obtained using a  sublinear asymptotic behaviour for $\gk$.

 We stress that our main purpose is to test whether or not $\gk=o(\bt)$ as $\bt\to\infty$ is a suitable asymptotic dependence for the non perturbative behavior of the Collins-Soper kernel. In this sense, \eref{gk-largeb-jogh} and \eref{gk-largeb-vlad} should be seen only as a proxy for such hypothesis. Consideration of two models for $\gk$ will allow us to get a ``measure'' of the correlations between $\md$ and $\gk$ and of the theoretical uncertainties introduced by model choices.

 \subsection{Behavior of $\gk$ in the small-$\bt$ limit.\label{s.small-bt-gk}}

 There is a general consensus that the behavior of $\gk$ in the small-$\bt$ limit should be power-like, see for example Refs.~\cite{Collins:2014jpa, Aidala:2014hva, Collins:2017oxh, Vladimirov:2020umg,Scimemi:2019cmh,Bacchetta:2019sam}. Often phenomenological studies have assumed 
 \begin{align}
\gk\overset{\bt\to0}{\sim} c_1 \bt^2 \,.
\label{e.gksmall}
\end{align}
For instance, Ref.~\cite{Bacchetta:2019sam} uses 
 \begin{align}
\gk = c_1 \bt^2 + c_2 \bt^4\,,
\end{align}
 where a strong suppression at small $\bt$ was necessary to reach a satisfactory description of Drell-Yan data at extremely large energies, which required high accuracy in the perturbative and  logarithmic expansion.
For this analysis, where the perturbative expansion only extends to NLL, 
we start by testing two different models for $\gk$ 
which ensure a $\bt^2$ behaviour at small $\bt$, while respecting the asymptotic trends discussed above. More specifically, we look at the following functional forms:
\begin{align}
\label{e.gksmall21}
& c \log\left(1+\left(\mk\bt\right)^2\right)\,,\\
\label{e.gksmall22}
& a \, \bt ^{p_k} \, \Big(1 - e^{-b/a \; \bt ^{(2-p_k)}}\Big)\,.
\end{align}
Both models show some drawbacks. First of all, the parameter space is not well constrained. Moreover,  larger values of $\chi^2$ point to the inadequacy of the power $2$ for $\bt$.
In fact, in our preliminary tests we find that our fit is rather sensitive to the modulation of $\gk$ in the large $\bt$ region.
Remarkably, it shows a strong preference for a sub-linear power or logarithmic raise of $\gk$, 
while definitely ruling out the 
$\bt^2$ or $\bt^4$ 
behaviour at large $\bt$. Indeed, it is likely that increased perturbative accuracy could accommodate for the behaviour of \eref{gksmall} at small $\bt$.
We therefore relax the constraint that $\gk$ should go to zero quadratically in the small $\bt$ limit, by simply requiring it to go to zero as some generic power $\pk>0$. This will also allow us to reduce the number of free parameters for our final analysis. Thus, we will focus on the following parametrizations
\begin{align}
\label{e.gk-jogh}
\gk &= \log\left(1+(\mk\bt)^{\pk}\right)\,\\
\label{e.gk-vlad}
\gk &= (\mk \bt)^{(1-2\pk)},\qquad 0<\pk<1/2\,.
\end{align}

The functional forms in \eref{gk-jogh} and \eref{gk-vlad}, labelled $A$ and $B$ respectively, are summarized in \tref{models3}. They optimize the quality of the fit while keeping the number of free parameters under control.

\bigskip

\subsection{Final Models and Data Kinematics.}
\label{s.finalmodelsandkin}
For our main analysis we focus on the following kinematics
\begin{align}\label{e.kin1}
0.375\leq\zh\leq 0.725\,,\qquad0.750\leq T\leq 0.875\,,
\end{align}
corresponding to Region 2 (see Ref~\cite{Boglione:2021wov}). 
Furthermore, as the TMD formalism of Ref.~\cite{Boglione:2020auc,Boglione:2021wov} regards the region in which $\qt=\pt/\zh\ll Q$,  we adopt the cut
\begin{align}\label{e.kin2}
\qt/Q\leq 0.15\,,
\end{align}
which gives us some confidence that the appropriate collinear-TMD factorization theorem is applied,
and perform a standard $\chi^2$ minimization procedure for each one of the models summarized in \tref{models3}. 
More restrictive cuts make it difficult to find an optimal solution, while less stringent ones result in large values of $\chi^2$.

As mentioned before, for our analysis  we consider two different models for each $\md$ and $\gk$, in order to provide a reliable estimation of the uncertainties affecting the extraction of the TMD FF. For $\gk$ we consider the functional forms in \eref{gk-jogh} and \eref{gk-vlad}, which we call model A and model B respectively. For $\md$, our starting point is the Fourier transform of a power law in momentum space, taking into account that a $z_h$-dependence is necessary for a successful description of the BELLE cross sections \cite{Seidl:2019jei}. These two models, labelled I and II, differ only in the treatment of the $\zh$ dependence. In total we have four different cases we will use, which we label as models IA, IB, IIA, and IIB.

\subsubsection{Models IA and IB}
\label{s.models1}

Model I for $\md$ was already introduced in \sref{largebT} (see \tref{models2}) and, 
as summarized in \tref{models3}, it concentrates the full $\zh$ dependence of $\md$ within the extra $F(\bt,\zh)$ factor, which is controlled by the mass-parameter $M_z = -M_1 \log (\zh)$, while the Bessel function and other factors corresponding to the power law in momentum space only depend on $\bt$. 

Thus, models IA and IB have initially six parameters each. In both cases, we find that when trying to fit all of the parameters simultaneously, some are poorly constrained and/or show very strong correlations. This may indicate some "redundancy", i.e. the existence of non-independent parameters. This can be an issue when attempting to provide a transparent statistical interpretation of results. We find that we have to fix a total of three parameters, two for $\md$ and one for $\gk$ in order to avoid such situation. 
We choose to fix the dimensionless powers, $p$, $q$ and $\pk$, so that we will find best fit values of parameters that  may have the interpretation of a "typical mass" of the observables. 
\begin{figure*}
		\centering
		\subfloat[]{\includegraphics[scale=1]{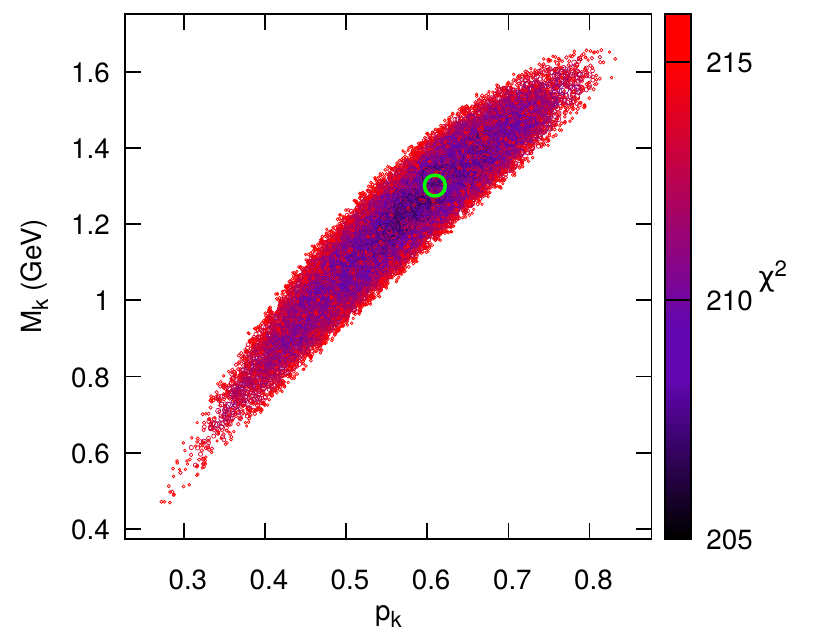}}
		\subfloat[]{\includegraphics[scale=1]{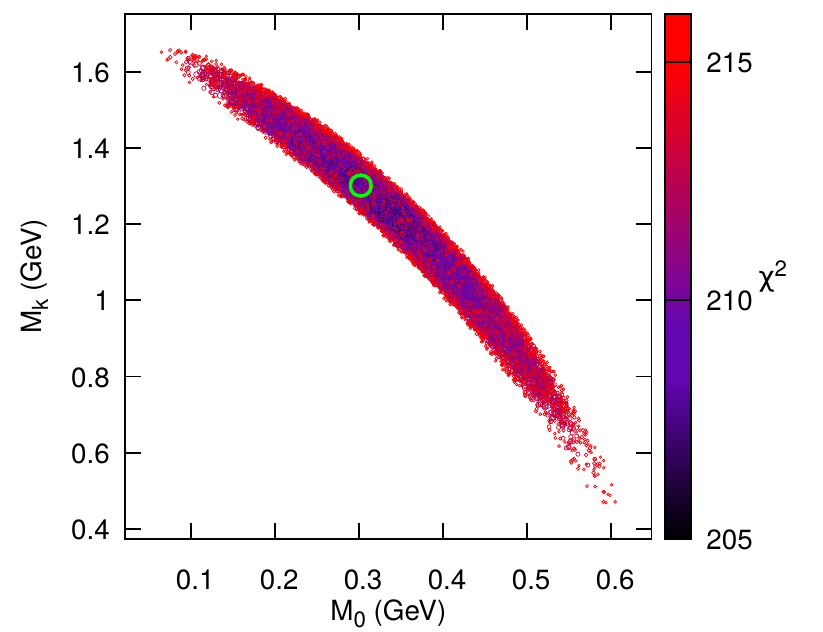}}
		\caption{
		Preliminary study of parameter space using model I for $\md$ and A for $\gk$, see  \tref{models3}, with fixed $p=1.51$ and $q=8$. The circles represent  parameter  configurations  in a region  where a minimum is found. The empty circles display the value of $\chi^2$ both by color (as in palette) and size (larger circles for smaller values of $\chi^2$), for configurations with $\chi^2_i<\chi^2_0+\Delta\chi^2$, with $\Delta\chi^2=9.72$. This value of $\Delta\chi^2$ corresponds to a $2\sigma$ confidence level for varying 4 parameters simultaneously, in situations where the $\chi^2$ as a function of parameters can be approximated as an ellipsoid around the minimum. In this case, however,  such approximation is not valid, hindering an interpretation in terms of confidence levels. Strong correlations as those shown likely indicate some "redundancy" in parameter space. (a) Correlation between $\mk$ and $\pk$, where the green circle indicates the minimal configuration. (b) Correlation between $\mk$ and $\mo$.
		\label{f.corr1}
	}
\end{figure*}
First, we set $p=1.51$, so that  the derivative of the Bessel function in model I vanishes at $\bt=0$, this prevents $\md$ from being sharply peaked at $\bt=0$. After setting the value for $p$, we find that the minimum\footnote{More precisely a "lower bound", not the minimum $\chi^2$ in the mathematical sense. } value of $\chi^2$ one can obtain (for both models IA and IB), corresponds to $q\approx8$, so we fix $q=8$. Finally, provided this choices for $p$ and $q$, we perform a fit in order to obtain the optimal values for the power parameter $\pk$ for each model IA and IB. We show the results of this last step in \fref{corr1} for model IA (fixed $p=1.51$, $q=8$ and varying $\pk$), in order to illustrate the need to fix some of the parameters. There, the circles display parameter configurations $i$ with $\chi^2_i$ values that deviate from the minimum $\chi^2_0$ by no more than a ``tolerance''\footnote{This value corresponds to a 2$\sigma$ confidence level for varying 4 parameters, but we do not attempt to make such an interpretation in this particular case.} $\Delta \chi^2=9.72$; green dots represent the minimal configuration.
While in this case it is possible to find a minimum varying $\mo$, $M_1$, $\mk$ and $\pk$ simultaneously, 
 very strong correlations appear and parameter configurations  significantly deviate from ellipsoidal shapes , as shown in \fref{corr1}. This makes it difficult to draw regions in parameter space as it is  usually done, by considering  configurations for which $\chi^2_i<\chi^2_0+\Delta\chi^2$ \emph{and} interpret them in terms of confidence levels,  i.e. statistical errors of our analysis. 
 As we will see in the next section,  by varying only the three mass parameters $\mo$, $\mz$ and $\mk$, parameter space display elliptical profiles for all correlations, allowing for a more sound statistical interpretation.
 It is interesting to note that the strong correlations appear also between $\md$ and $\gk$ parameters, as seen in the right panel of \fref{corr1}. 

The information regarding the values of $p$, $q$ and $\pk$ is summarized summarized in \tref{models3}. 
We remark that these choices 
still allow for enough flexibility in our models.

Note that while we could have treated $\pk$  as nuisance parameters, for our purposes it is enough to fix them to reasonable values, since we are mostly interested in addressing the compatibility of the asymptotic behaviour of \eref{modelasyconst}, \eref{gk-largeb-jogh} and \eref{gk-largeb-vlad} with BELLE data; for this, it is suffices to consider reasonable profile functions. A possible concern regards the estimation of statistical errors, which may be affected  by fixing parameters. However, we 
remark that considering different models helps us in giving an estimate of some of the theoretical  uncertainties of our extraction. All of our choices for models IA and IB are summarized in \tref{finalchi21}.

\subsubsection{Models IIA and IIB}

Model II stems from different considerations, namely, we do not introduce the extra factor $F$ but rather assign a $\zh$ dependence to the mass and power parameters of the Bessel function themselves, $M$ and $p$. 
This offers a nice physical interpretation, especially if we recall that this $\bt$-distribution originates as the Fourier transform of a
power law, which resembles a propagator, 
of the form
$[M(z)^2 + q_T^2]^{-p(z)}$ in $\qt$-conjugate space.
In this sense, the mass $M(z)$ can be regarded as an \emph{effective mass}, that modifies the mass of the detected hadron $M_h$ in a $z_h$-dependent way. The power $p(z)$ can be re-written as $p(z) = 2 + \gamma_P(z)$, where the whole $z_h$-dependence has been encoded into an 
\emph{anomalous dimension} $\gamma_P$.
As for model I, the strong correlations between $p(z)$ and $M(z)$  makes it impossible to extract them simultaneously in a converging  fit: therefore, further constraints are required to be able to proceed with our analysis. 
For model II we constrain the $\zh$ behavior of $\md$ by analytically requiring that the theory lines appropriately reproduce some basic features of the measured cross section, namely the peak height and the width of the $P_T$ distributions, at each single measured value of the kinematic variable $\zh$.
In particular~\cite{Seidl:2019jei}, the width of the measured cross section reaches its maximum at intermediate values of $z_h$ (around $\sim 0.6$, as obtained in Ref.~\cite{Seidl:2019jei}) for all thrust bins belonging to the $2$-jet region. This property  can be used as a constraint for the model with the help of a proper change of variables, that trades $p$ and $M$ for the width $W$ and the peak height $P$ 
\begin{align}
    \label{eq:PandW}
    &p = \frac{1}{2} \left(\frac{3}{1-R}-1\right);
    &\quad M = \frac{W}{z} \, \sqrt{\frac{3}{1-R}}\,,
\end{align}
where $W \geq 0$ and $R$  is the ratio  ${P}/{P_{\text{\small max}}}$ between the peak height and its maximum possible value ($0 < R < 1$).
The advantage of this operation is that $R$ and $W$ can be regarded as variables associated to the full cross section and not only to the TMD model. 
However, being a mere change of variables, this does not solve any correlation issues, which are simply being moved from ($p$, $M$) to ($R$, $W$). 
In particular, observation shows that $R$ and $W$ are inversely proportional with respect to their $z_h$-dependence: where one shows a maximum the other has a minimum and vice-versa. 
Therefore, we set:
\begin{align}
    \label{eq:PandW_1ansatz}
    &R = f(z_h,z_0);
    &\quad W = \frac{M_\pi}{f(z_h,z_0)^2},
\end{align}
where $M_\pi = 0.14$ GeV is the mass of charged pions and $f$ has to be a positive-definite function, never larger than $1$ and with a minimum in $z_h=z_0$. This is where the information associated to the experimental observation comes into play, 
helping to select an appropriate $z_h$-dependence for the TMD model. In fact, 
the function $f$ has a minimum 
in the exact point where the width $W$ has a maximum.
One of the simplest functional forms which fulfills such requirements is 
\begin{align}
    \label{eq:f_def}
    f(z,z_0) = 1 - (1-z)^\beta,
    \quad\text{with }\beta = \frac{1-z_0}{z_0}.
\end{align}
This is what we adopt for Model II. 
The expression of $M_z$ and $p_z$ in terms of $f(z)$ 
are summarized in \tref{models3}.

\begin{table}
	\caption{Models for $\md$ and $\gk$ in impact parameter space for our main analysis. $\md$ is obtained by multiplying the BK model, which corresponds to a power law in momentum space, with an additional function of $\bt$ and $\zh$. 
	}
	\label{t.models3}
	\begin{center}
		\begin{tabular}{|c|l|c|}
			\hline
			\multicolumn{3}{|c|}{\multirow{2}{*}{}}\\
			\multicolumn{3}{|l|}{\multirow{2}{*}{$\qquad\md=\cfrac{2^{2-p} (\bt \mo)^{p-1} }{\Gamma (p-1)}K_{p-1}(\bt \mo)\,\,\times\,\,F(\bt,\zh)$}}\\
			\multicolumn{3}{|c|}{\multirow{2}{*}{}}\\
			\multicolumn{3}{|c|}{\multirow{2}{*}{}}\\
			\hline
		    \hline
			ID&$\qquad\qquad \md$  model&parameters\\
			\hline
			\multirow{3}{*}{ I }&\multirow{3}{*}{
				$F=\left(\cfrac{1+\log\left(1+(\bt M_z)^2\right)}{1+(\bt M_z)^2}\right)^q$}&\multirow{2}{*}{$\mo,\,\mz$
			}\\
			& & \\
			& & $p=1.51,\,\,q=8$ \\
			\multirow{2}{*}{ }&\multirow{2}{*}{$\,M_z=-\mz\log(\zh)$}& 
			\\
			& & \\
			\hline
			\multirow{2}{*}{ II }&\multirow{2}{*}{$F=1$}& \\
			& &$z_0$ \\
			\multirow{3}{*}{ }&\multirow{3}{*}{$\,
			M_z = M_h \, \cfrac{1}{z \, f(z)^2} \, \sqrt{\cfrac{3}{1-f(z)}}
			$ } & 
			\\
			& & \\
			& & \\
			\multirow{3}{*}{ }&\multirow{3}{*}{$\,p_z = 1 + \cfrac{3}{2}\;\cfrac{f(z)}{1-f(z)}$} &
			\\
			& & \\
			& & \\
			\multirow{3}{*}{ }&\multirow{3}{*}{$f(z) = 1 - (1-z)^\beta$, ~ $\beta = \frac{1-z_0}{z_0}$ } 
			& 
			\\
			& &  \\
			& & \\
			\hline
			\hline
			\multicolumn{3}{|c|}{\multirow{1}{*}{$\gk$ model}}\\
			\hline
			\multirow{3}{*}{A}&\multirow{3}{*}{$\,\,\,\gk=\log\left(1+(\bt \mk)^{\pk}\right)$}&\multirow{3}{*}{$\mk,\,\,\,\pk$}\\
			& & \\
			& & \\
			\hline
			\multirow{3}{*}{B}&\multirow{3}{*}{$\,\,\,\gk=\mk \bt^{(1-2\pk)}$}&\multirow{3}{*}{$\mk,\,\,\,\pk$}\\
			& & \\
			& & \\
			\hline
		\end{tabular}
	\end{center}
\end{table}

Following the indication of these preliminary tests, we will focus on the study of the large $\bt$ (i.e. small $P_T$) behaviour of the fitted cross sections, leaving the exploration of the small $\bt$ region to further analyses. 
By large $\bt$, here we mean ``the largest $\bt$ experimentally accesible'', as the asymptotic behaviour may not be so relevant for this data set, as discussed in \sref{largebT}. 
For our main analysis with model II,  we will  adopt the functional forms of \eref{gk-jogh} and \eref{gk-vlad}, both characterized by two free parameters, $\mk$ and $\pk$. This gives two new models, which we label ``IIA'' and ``IIB'' (see \tref{models3}).

 We thus minimize $\chi^2$ with respect to the free parameters ($z_0$, $\mk$, $\pk$) for models IIA and IIB. 
In these two cases, as for model I, we will estimate statistical errors by determining the 2$\sigma$ confidence region in parameter space. Note that, while parameter space shown in next section for model II has a distortion respect to elliptical shapes, we have checked that rescaling the parameters allows to correct for this. Nonetheless, we present results in terms of ($z_0$, $\mk$, $\pk$) since they are closely related to features of the data.

Following the above considerations, the main results of our analysis will be presented in the next  subsection for all of our models.

\subsection{Phenomenological Results.\label{s.pheno-res} }

With our final choices, we perform  fits for each of the considered models, 
labeled  IA, IB, IIA, IIB, where "I" and "II" indicate the choice of parametrization for $\md$ while "A" and "B" indicate  the model chosen for $\gk$, according to the notation introduced in \tref{models3}. In each case we perform a $\chi^2$-minimization procedure using MINUIT \cite{James:1975dr}, fitting a total of 3 parameters in each model. 
We estimate parameter errors by considering 2$\sigma$ confidence regions. 
In other words, for each model we  consider  configurations in parameter space around the minimal one, varying all parameters simultaneously and
accepting those for which $\chi^2_i<\chi^2_0+\Delta\chi^2$, with $\Delta\chi^2=8.02$; this value of $\Delta\chi^2$ is consistent with varying three parameters simultaneously. Final results for models IA and IB are reported in \tref{finalchi21}. For models IIA and IIB, results are displayed in \tref{cut1_andrea}.

From a  superficial look at \tref{finalchi21}, one may  conclude that the quality of model IB is higher, given the smaller values of $\chid$.
However, we note that model IB has the disadvantage that the ellipsoidal approximation extends down to negative values of $\mo$, which must be excluded. This is reflected by the asymmetric errors in $\mo$ and $\mk$ in the third column of \tref{finalchi21}. 
%
%

\begin{table}[t]
\caption{
\label{t.finalchi21}
Minimal $\chid$ obtained by fitting models IA and IB, 
according to table \tref{models3}. In each case we perform fits in the kinematical 
region of \eref{kin1} and \eref{kin2}. 
In both cases IA and IB, all dimensionless parameters are fixed, indicated by in the table by a star. Fixed values as explained in \sref{finalmodelsandkin}. }

\begin{center}
\begin{tabular}{c c c }
\hline
\multicolumn{3}{c}{$\qt/Q<0.15$ ($\text{pts}=168$)}\\
\hline
&IA&IB \\
\hline
\rows{$\chid$}
&\rows{$1.25$}
&\rows{$1.19$}
\\
& & \\
\rows{$\mo(\gev)$}
&\rows{$0.300^{+0.075}_{-0.062}$}
&\rows{$0.003^{+0.089}_{-0.003}$}
\\
& & \\
\rows{$\mz(\gev)$}
&\rows{$0.522^{+0.037}_{-0.041}$}
&\rows{$0.520^{+0.027}_{-0.040}$}
\\
& & \\
\rows{$p^{*}$}
&\rows{$1.51$}
&\rows{$1.51$}
\\
& & \\
\rows{$q^{*}$}
&\rows{$8$}
&\rows{$8$}
\\
& & \\
\rows{$\mk(\gev)$}
&\rows{$1.305^{+0.139}_{-0.146}$}
&\rows{$0.904^{+0.037}_{-0.086}$}
\\
& & \\
\rows{$\pk^{*}$}
&\rows{$0.609$}
&\rows{$0.229$}
\\
& & \\

\hline
\end{tabular}
\end{center}
\end{table}

Fits performed with model II have slightly higher $\chi ^2$s, as shown in \tref{cut1_andrea}. This is  probably due to the fact that this model, being more tightly constrained, with only one free parameter controlling the $\zh$ behaviour of $\md$, shows a limited flexibility compared to model I.
Nonetheless clear differences between models cannot be observed when comparing to data. 
We thus consider both models I and II as equally acceptable to describe the general profile of our functions $\md$ and $\gk$.
We choose model IA to display the agreement of our predicted cross sections to the BELLE data in \fref{fitdata}, noting that corresponding comparisons for models IB, IIA, IIB would indeed be  very similar. \fref{fitdata} shows two types of errors bands.  Darker colored bands represent the  statistical uncertainty of the fit. The lighter colored bands are an estimate of the  error induced by the collinear fragmentation functions used in the analysis. They are produced by refitting the model function for each of the replicas provided by the NNFFs NLO extraction of Refs.~\cite{Bertone:2017tyb}

 For this estimate, only about $65\%$ of the NNFFs replicas allowed for a convergent fit. A more detailed study of such errors is  a necessity in this type of studies that need constraints from independent analyses. For now, we consider our estimate as a useful tool to understand the effect of the choice of collinear FFs in a TMD extraction. In fact, it is 
 useful to observe in \fref{fitdata} that errors from the collinear functions are consistently larger than statistical errors. Arguably, the former render a more realistic picture of the precision at which TMDs can be  extracted from data. 
 It is clear from \fref{fitdata} that the quality of the description of data deteriorates at smaller values of $T$. This is not surprising since the formalism employed \cite{Boglione:2020cwn,Boglione:2020auc,Boglione:2021wov} is expected to fail at smaller values of thrust, where the topology of the $\epm \to hX$ events starts deviating from a 2-jet like  configuration. 
 %

\begin{table}[t]
\caption{Minimal $\chid$ obtained by fitting models IIA and IIB, 
according to table \tref{models3}. In each case we perform fits in the kinematical 
region of \eref{kin1} and \eref{kin2}. There are no nuisance  parameters in model II.}
\label{t.cut1_andrea}
\begin{center}
\begin{tabular}{c c c }
\hline
\multicolumn{3}{c}{$\qt/Q<0.15$ ($\text{pts}=168$)}\\
\hline
%
%
%
&IIA&IIB \\
\hline
\rows{$\chid$}
&\rows{$1.35$}
&\rows{$1.33$}
\\
& & \\
\rows{$z_0$}
&\rows{$0.574^{+0.039}_{-0.041}$}
&\rows{$0.556^{+0.047}_{-0.051}$}
\\
& & \\
\rows{$\mk(\gev)$}
&\rows{$1.633^{+0.103}_{-0.105}$}
&\rows{$0.687^{+0.114}_{-0.171}$}
\\
& & \\
\rows{$p_k$}
&\rows{$0.588^{+0.127}_{-0.141}$}
&\rows{$0.293^{+0.047}_{-0.038}$}
\\
& & \\

\hline
\end{tabular}
\end{center}
\end{table}

\begin{figure*}
	\centering
	\includegraphics[scale=1]{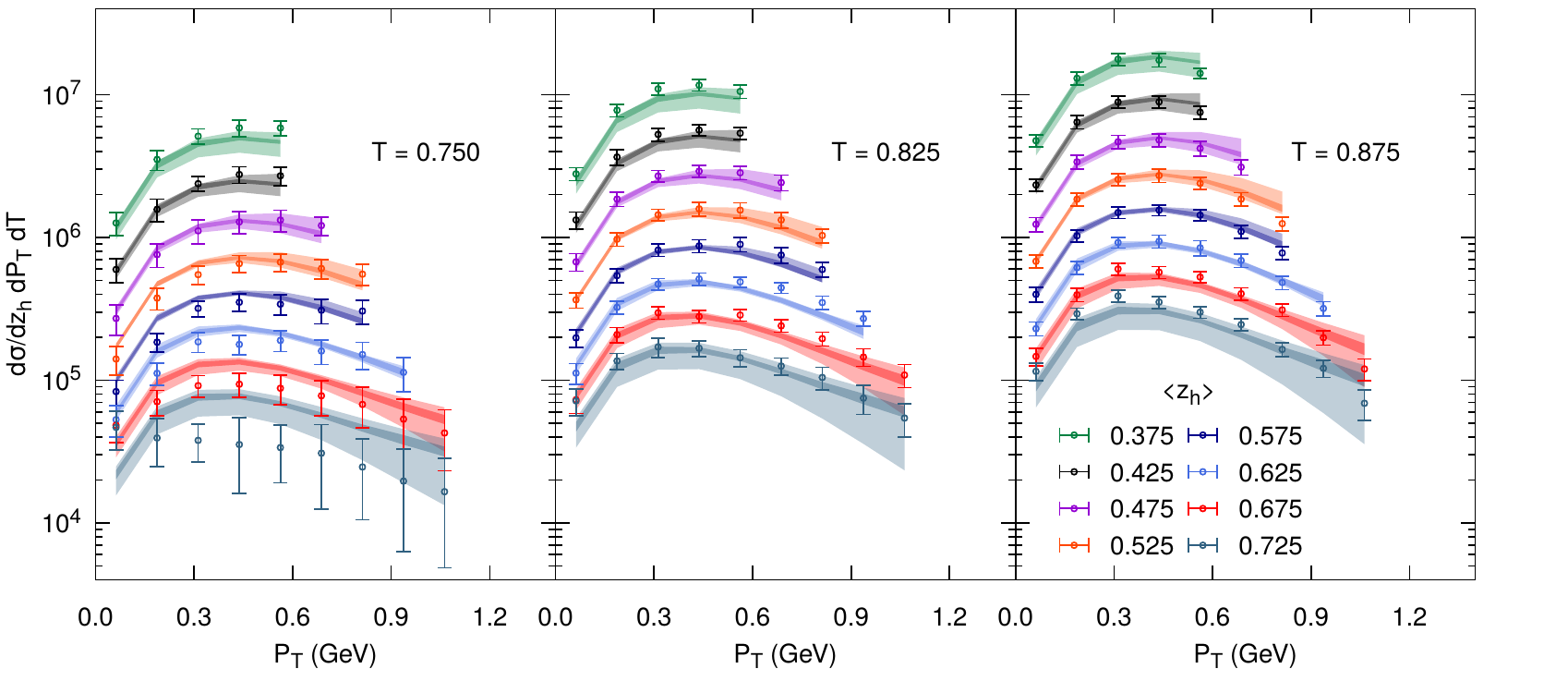}
	\caption{
		Results of fitting model IA from \tref{models3}, in the kinematical  region of \eref{kin1} and \eref{kin2}. Darker shaded bands represent the  statistical uncertainty of the fit at 2$\sigma$ confidence level, and correspond to the  parameter configurations of \fref{corr3}. The lighter shaded bands are an estimate of the  error induced by the collinear fragmentation functions used in the analysis, and are produced by refitting the model function for each of the replicas provided by the NNFFs NLO extraction of \cite{Bertone:2017tyb}. For a better visualization of results, central lines are not included, but they generally lie in the middle of the thin, darker statistical error bands. Models IB, IIA, IIB give  analogous results. We do not show them in the plot as they would be indistinguishable.
		\label{f.fitdata}
	}
\end{figure*}
%
Further developments in the theoretical treatment of the interplay between the rapidity divergence regularization and the thrust dependence 
will likely improve the quality of the extraction by allowing the possible inclusion of more data points while achieving an improved agreement to data~\cite{Boglione:2021wov}.
We leave this for future work~\cite{Boglione-Simonelli:2022}.

Interesting results are found about $\gk (b_T)$. We focus on the study of the large $\bt$ (i.e. small $P_T$) behaviour of the fitted cross sections, leaving to further analyses the exploration of the small $\bt$ region, on which we are unable to draw definite conclusions, as explained in \sref{small-bt-gk}.
Our fit is rather sensitive to the modulation of $\gk$ in the large $\bt$ region. Remarkably, it shows a strong preference for a sub-linear power or logarithmic raise of $\gk$, while definitely ruling out the 
$\bt^2$ or $\bt^4$ behaviour at large $\bt$.
We stress that 
by large $\bt$, here we mean ``the largest $\bt$ experimentally accesible'', as the asymptotic behaviour may not be so relevant for this data set, as discussed in \sref{largebT}.

 It is important to understand the strength of  correlations between $\md$ and $\gk$ and the impact of  model choices in the extraction of profile functions. Although these two points are not necessarily unrelated, we discuss them separately in what follows.  

	\begin{figure*}
		\centering
		\subfloat[]{\includegraphics[scale=1]{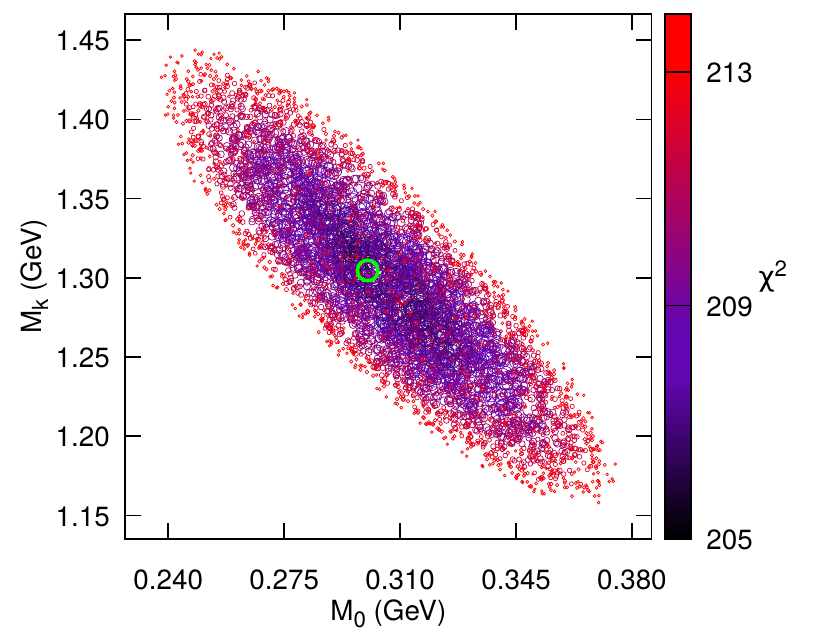}}
		\subfloat[]{\includegraphics[scale=1]{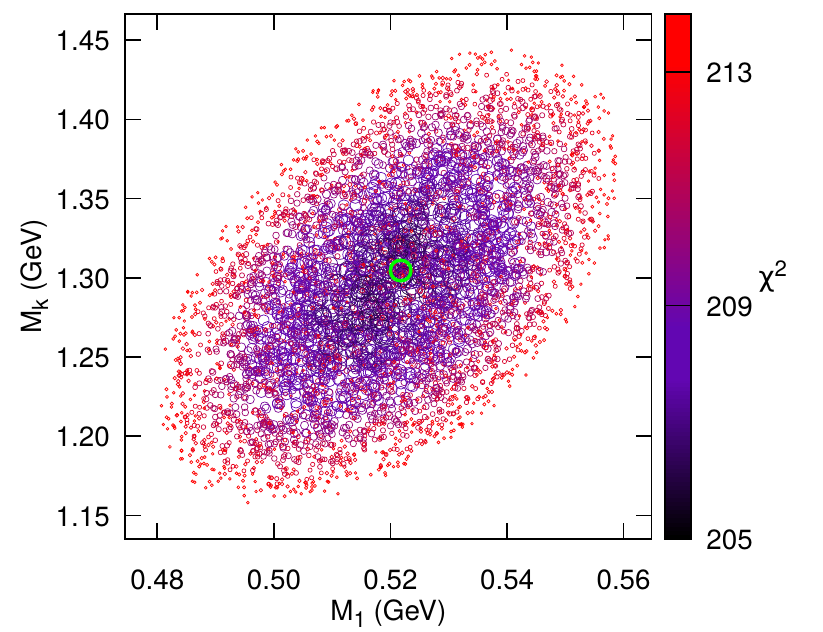}}
		\caption{
			\label{f.corr3}
			2$\sigma$ confidence regions centered around the minimum configuration, shown in green, for the fit of model IA of \ref{t.models3} in the kinematical region of \eref{kin1} and \eref{kin2}.
		}
	\end{figure*}

Firstly, regarding correlations between $\md$ and $\gk$ for a given model,  in an ideal scenario one would expect them to be mild, which would provide some level of confidence when comparing results to other analyses or data sets. This situation is however not guaranteed. We find that in fact $\md$ and $\gk$ are 
correlated, as shown in \fref{corr3}, where correlations  between $\mk$ and the mass parameters of $\md$, $\mo$ and $\mz$ are displayed for model IA, and in \fref{corr4} where analogous scatter plots are presented for model IIB, for the correlation of $z_0$ with $\mk$ and $\pk$. We obtain analogous results for model IB, with the added feature that confidence regions in parameter space appear as  ellipses truncated in the region $\mo<0$. 
For models of type II, the correlation between $\md$ and $\gk$ appears to be stronger than in the parametrizations of type I, so much so that a slight residual deformation from the ellipsoidal form is still visible in \fref{corr4}, although the constraints intrinsically built in model I drastically limit the number of its free parameters. 
We checked that a transformation of parameters $\mk$ and $\pk$ render scatter plots 
with an approximate elliptical shape.
It is noteworthy, 
that the regions corresponding to 2$\sigma$ confidence level have  well defined contours, allowing for a reliable determination of the error affecting the extracted parameters.

Secondly,  we find that the profile of the extracted functions strongly depends on model choices.
Note that the full TMD in momentum space, shown in \fref{tmdu}, shows differences beyond statistical error bands. Discrepancies are more visible when considering separately the results obtained for the extractions of $\md$ and $\gk$, as seen in \fref{mdktilde} where the 
profile functions 
differ beyond  statistical error bands. As such, those discrepancies should be considered as a kind of theoretical error. 
While this is only a rough estimate of one kind of theoretical uncertainties, it makes the case that statistical uncertainties are generally not enough to asses the quality of an extraction. Even though this is specially the case in studies like  the present one, where only one process is considered, it is a matter of concern even for global fits. 
\begin{figure*}
		\centering
		\subfloat[]{\includegraphics[scale=1]{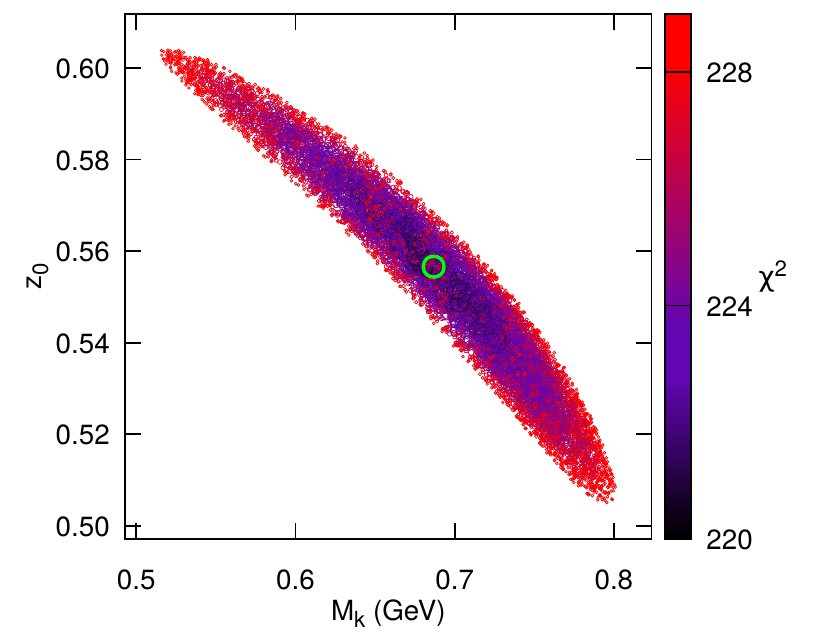}}
		\subfloat[]{\includegraphics[scale=1]{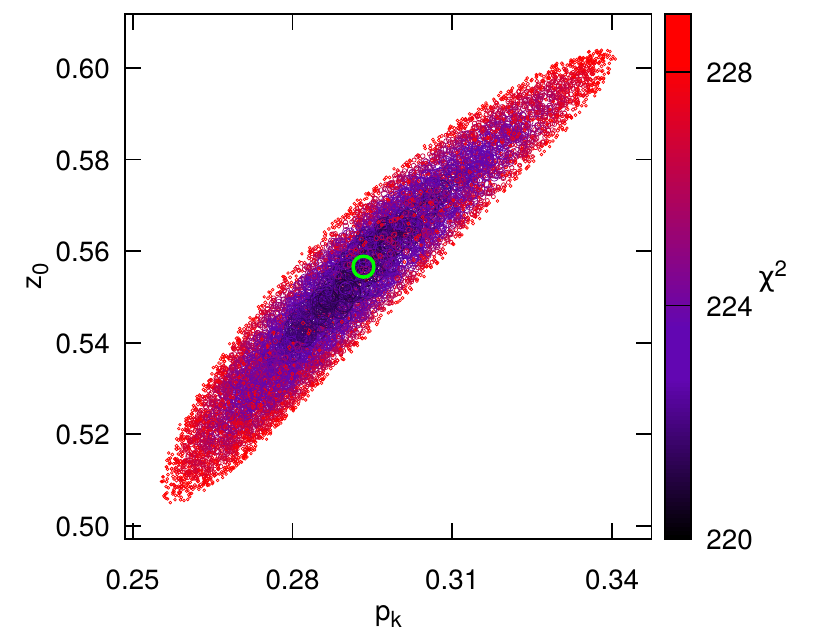}}
		\caption{
			\label{f.corr4}
			2$\sigma$ confidence regions centered around the minimum configuration, shown in green, for the fit of model IIB of \ref{t.models3} in the kinematic region of \eref{kin1} and \eref{kin2}. Here the presence of some correlation among the free parameters controlling the behavior of $\md$ and $\gk$ is signalled by a slight deformation from the expected ellipsoidal shapes.
		}
		\end{figure*}

\begin{figure}[ht]
	\centering
	\includegraphics{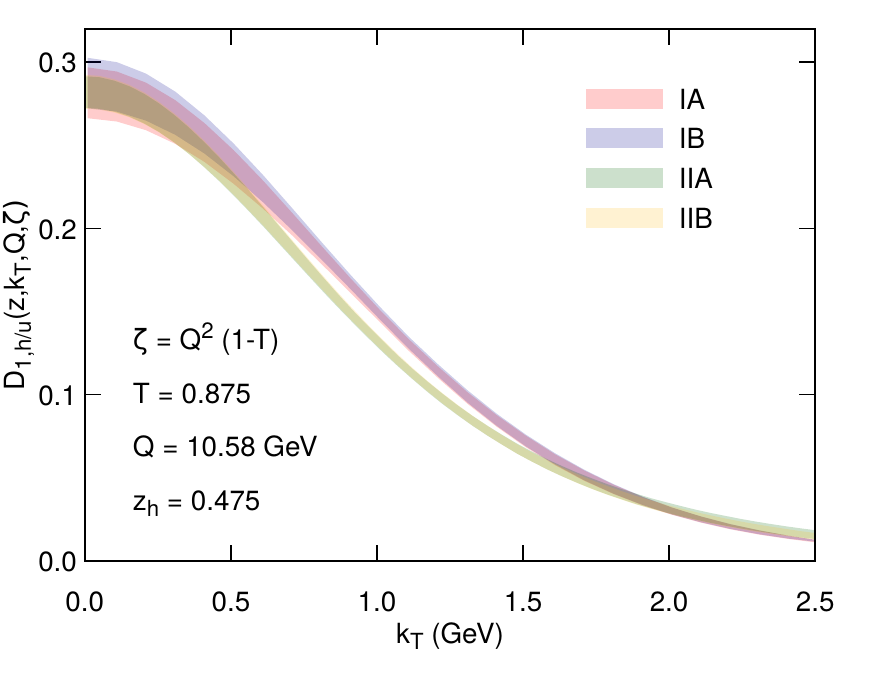}
	\caption{
		Extractions of the unpolarized TMD FF, \eref{tmd_NLL}, from one-hadron production BELLE data of \cite{Seidl:2019jei}, using models IA,IB,IIA,IIB of \tref{models3}, in the kinematic region of \eref{kin1} and \eref{kin2}. The TMD FF for the $u \to \pi^+ + \pi^-$ channel is shown in momentum space.
		\label{f.tmdu}
	}
\end{figure}

\begin{figure*}
	\centering
	\includegraphics[]{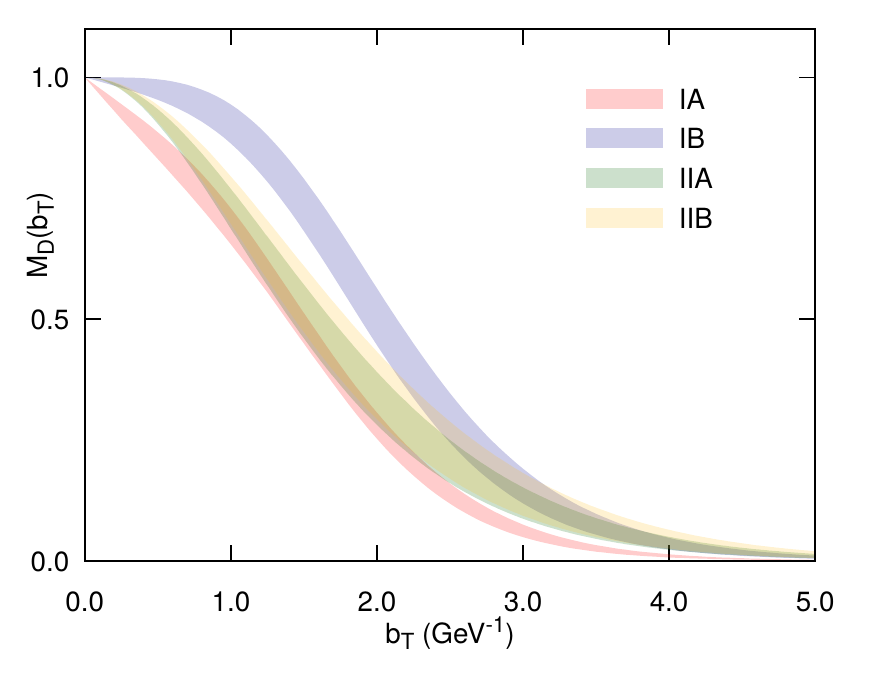}
	\includegraphics[]{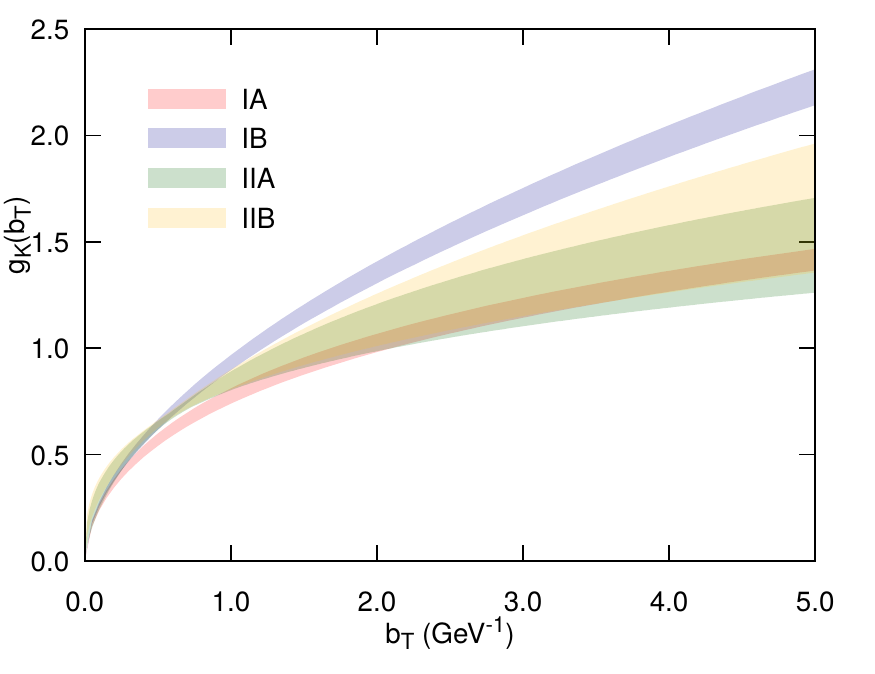}
	\caption{ 
		Extractions of $\md$ and $\gk$ in \eref{tmd_NLL} from $\epm \to hX$ BELLE data \cite{Seidl:2019jei}, in the kinematic region of \eref{kin1} and \eref{kin2}. In all cases, 2$\sigma$ statistical error bands are shown. For model IA they  correspond to the region of parameter space of \fref{corr3} while for model IIB to \fref{corr4}. Left: $\md$ according to model IA,IB,IIA,IIB of \tref{models3}. Right: Corresponding results for $\gk$.
		\label{f.mdktilde}
	}
\end{figure*}
We now compare our results against other recent TMD-analyses.
Since the relevant TMD FF in our studies is  different from that of the usual CSS, SCET and related treatments (see \eref{sqrtMD}), we can only compare our results for the CS kernel which, up to trivial constant factors, is the same in each scheme. 
In \fref{vlad} we plot the CS kernel~\cite{Collins:2011zzd,Collins:2017oxh} computed to NLL-accuracy
 \begin{align}
      \tilde{K}(\bt;\mu)&= \frac{1}{2}\Bigg[
     g_1^{\text{K}} (\lambda) + \frac{1}{L_b^\star} \, g_2^{\text{K}}(\lambda)\bigg] 
     - \frac{1}{2} g_K(b_T),
 \end{align}
 where the functions $g_1^{\text{K}}$ and $g_2^{\text{K}}$, which depend only on the combination $\lambda = 2\,\beta_0 \, a_S(\mu) \, L_b^\star$, with $L_b^\star = \log{\left( {\mu}/{\mu_{b_\star}}\right)}$, are reported in Appendix~\ref{app:tmd}.
%
Our extraction of the CS kernel for all our models is compared to the results obtained in the analyses of PV19~\cite{Bacchetta:2019sam} and SV19~\cite{Scimemi:2019cmh}\footnote{Note that for the CS kernel, PV19 follows the conventions of Ref.~\cite{Collins:2011zzd}, the SV19 results must be multiplied by a factor of $-2$ and ours should be divided by a factor 2.}.
For clarity, we don't show central lines but only error bands in each case.
\fref{vlad} shows a good agreement between our extraction of the  CS kernel  and the SV19 analysis in the region just above $\bt \sim 2$ GeV$^{-1}$. Note that these two extractions are based on different factorization schemes and exploit different data sets.
The large $\bt$ behaviour of our extraction is clearly different from the PV19 results, which adopts a $\bt ^4$ asymptotic behaviour in order to describe Drell-Yan production data from different experiments on a very wide kinematic range, and up to extremely high energies. 
%
Instead, in the small $\bt$ region, our extraction of the CS kernel  differs from both PV19 and SV19 results, 
where the perturbative part of the CS kernel is expected to dominate, making all bands to coincide.

This is  mostly due to two factors. First, the behaviour of our model for $\gk$ at small distances, which approaches zero only as $\bt^p$, with $0<p<1$, significantly more slowly compared to the $\bt^2$ behaviour of the PV19 and SV19 parametrizations also at small distances. In fact, the effects of our extractions for $\gk$ are still significant  at relatively small values of $\bt$. Second, the approximations of \eref{tmd_NLL}, are likely not optimal to describe the small $\bt$ behaviour of the TMDFF. Future improvements in the perturbative accuracy and a better treatment of the thrust dependence could resolve these discrepancies with respect to the results of the  PV19 and SV19 analyses. 

\begin{figure}
    \includegraphics{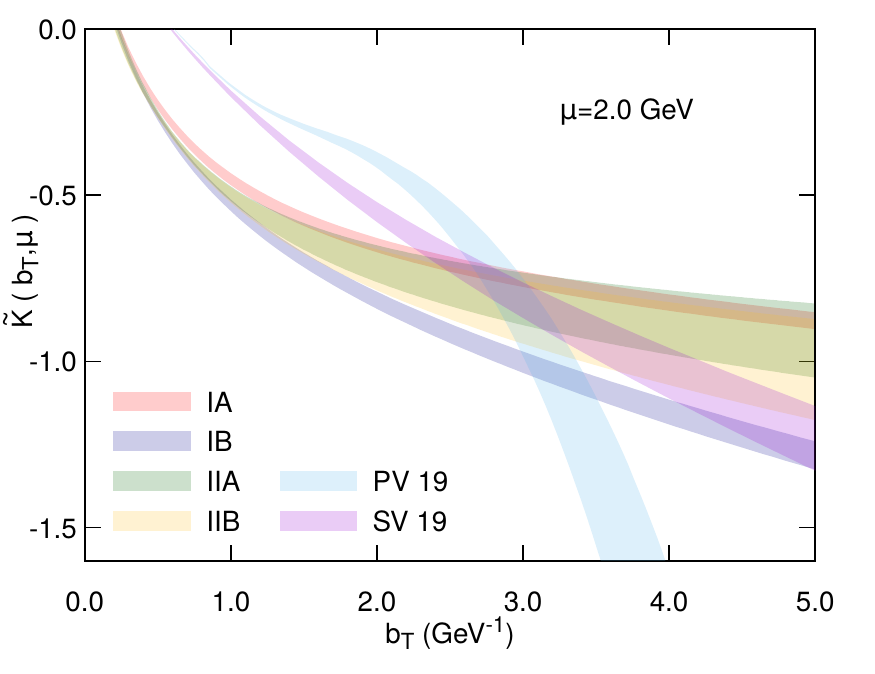}
	\caption{ Extractions of the CS kernel  obtained in this analysis with models IA, IB, IIA, IIB  are compared the PV19~\cite{Bacchetta:2019sam} and  SV19~\cite{Scimemi:2019cmh} extractions. For clarity, central lines are not shown.
	While there is a good agreement between the linear and sub-linear large $\bt$ behaviour 
	of this  extraction and Ref.~\cite{Scimemi:2019cmh}, the result of Ref.~\cite{Bacchetta:2019sam} shows an evident deviation at large $\bt$, where $\gk$  goes like $\bt ^4$. Discrepancies at small $\bt$ are due to the higher pQCD accuracy of the PV19 and SV19 analyses. We also note that our  models are essentially different  at small $\bt$ compared to those  used in Refs.~\cite{Bacchetta:2019sam,Scimemi:2019cmh}, as explained in the text.
	\label{f.vlad}
	}
\end{figure}

Recently, several lattice QCD  calculations of the CS kernel have been performed by different groups and reported in Refs.~\cite{Shanahan:2020zxr,latticeParton:2020uhz,Schlemmer:2021aij, Li:2021wvl,Shanahan:2021tst,LPC:2022ibr}; it is therefore interesting to compare our extraction to some of these computations. We do this in \fref{latticeCS}, where for clarity we compare error bands of all  our models 
with the most recent  calculation of each lattice QCD collaboration, 
Refs.~\cite{Schlemmer:2021aij, Li:2021wvl,Shanahan:2021tst,LPC:2022ibr}.
The logarithmic and sub-linear power large $\bt$ behaviour assumed for our extractions seem to be well supported by lattice QCD estimations of the CS kernel. We note that while our results are  in better agreement with the SWZ21\cite{Shanahan:2021tst} and LPC22\cite{LPC:2022ibr} calculations, the general trend of our extractions is also consistent with the ETMC/PKU\cite{Li:2021wvl} and SVZES\cite{Schlemmer:2021aij} results, characterized by a slow variation of the CS kernel at large $\bt$.
Once again we underline that in our analysis little can be said about the small $\bt$ behaviour of the CS kernel, thus we focused our attention in the large $\bt$ regime,  where BELLE experimental data offer good coverage.
\begin{figure}
\includegraphics{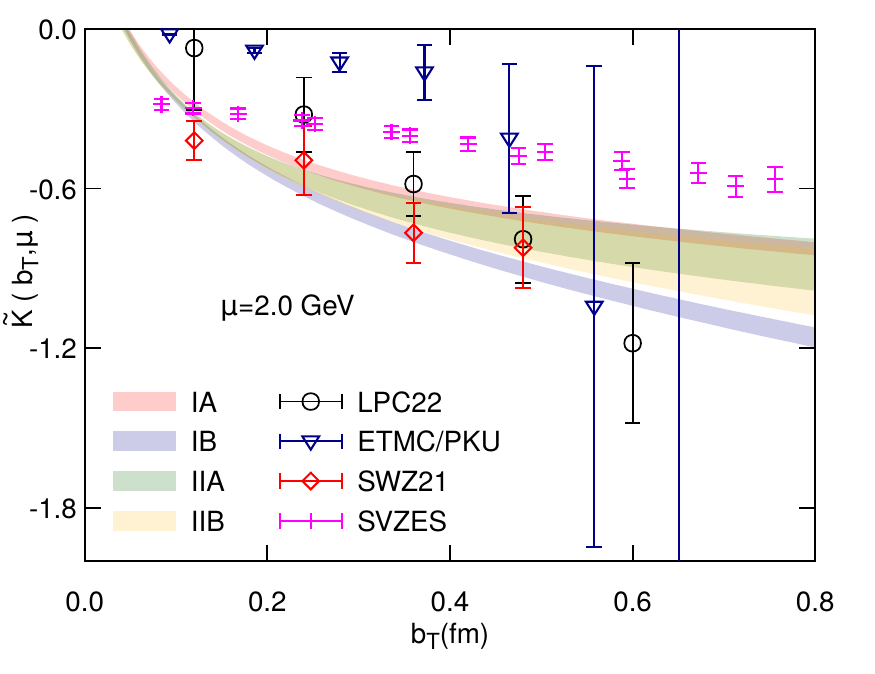}
	\caption{ The CS-kernel obtained in this analysis by adopting models IA, IB, IIA, IIB  are compared to the CS kernel  computed in lattice QCD in  Refs.~\cite{Schlemmer:2021aij, Li:2021wvl,Shanahan:2021tst,LPC:2022ibr}, at $\mu = 2$ GeV. For clarity, central lines for our extractions are not shown and we display only the most recent lattice calculation for each group.
	The logarithmic and sub-linear power large $\bt$ behaviour assumed for our extraction seem to be well supported by lattice QCD estimations of the CS kernel.
	\label{f.latticeCS}
	}
\end{figure}

\section{Conclusions}

 We performed an analysis of recent BELLE data  for one hadron production in $e^+e^-$ annihilation \cite{Seidl:2019jei} and extracted the TMD FF following the newly developed formalism of Ref.~\cite{Boglione:2020cwn,Boglione:2020auc,Boglione:2021wov}. In this framework, the short distance behavior of the TMD FF is constrained by collinear FFs, as in the standard CSS and SCET formalisms, while
the long distance behaviour requires the parametrization and determination, via comparison to data, of two functions, $\md$ and $\gk$. We  introduced constraints for these functions in the asymptotically large region of $\bt$, consistently with previous theoretical results from Refs.~\cite{Schweitzer:2012hh,Collins:2014jpa,Aidala:2014hva,Vladimirov:2020umg}. 
Our analysis is based on a maximum-likelihood procedure, carried out by $\chi^2$-minimization. Statistical errors are estimated by a standard determination of confidence regions at 2$\sigma$ level.

 Upon testing how different choices of available collinear FFs perform when comparing to data, we found that 
 both  JAM20~\cite{Moffat:2021dji} and  NNFF~\cite{Bertone:2017tyb} sets, although showing non-negligible differences (at least in some specific regions of $z_h$ and $b_{\text{T}}$)  are consistent with the $\pt$-dependent BELLE cross sections, within our approach.
 
 For our extraction,  constraints for both $\md$ and $\gk$  in the asymptotically large  $\bt$ region were imposed. For $\md$, we considered models characterized by an exponential asymptotic $\bt$ decay, according to previous theoretical  results from Ref.~\cite{Schweitzer:2012hh,Collins:2014jpa} and argued that, for consistency with universality of the large distance behavior of TMDs, the CS kernel should grow more weakly than a linear function of $\bt$ in the asymptotic limit. We considered two  models for $\gk$ satisfying that condition, which follow a sub-linear power and a logarithmic behavior, as suggested in 
 Refs.~\cite{Vladimirov:2020umg} and 
 \cite{Aidala:2014hva}, respectively, in this limit. We  showed that, in the considered  kinematic region, all aforementioned constraints imposed in the very large $\bt$ are consistent with the data. 
 We remark, however, that the asymptotic behavior of different models plays a role in extending results to smaller scales, and that  the slow evolution characteristic of the region of a few GeV can be accommodated by the type of models we tested in this work (see detailed discussion in Ref.~\cite{Collins:2014jpa}).
 
 A remarkable result of this analysis  is the insight of the influence of the profile function 
 of $\gk$ in the region of intermediate-moderate values of $\bt$, which we expect to be accessible at BELLE kinematics. Compared to previous studies~\cite{Bacchetta:2019sam, Scimemi:2019cmh}, which gave indications on the preferred behaviour of $\gk$ at small $\bt$, 
 our analysis based on the BELLE data, which correspond to a relatively moderate scale $Q = 10.6$ GeV, shows a significant sensitivity to  larger values of $\bt$.  We find clear signals that a $\bt ^2$ or $\bt ^4$ functional form is inappropriate to describe the long distance behaviour of the CSS kernel. In fact, the analyzed 
 data show a definite preference for a logarithmic or sub-linear modulation at large-$\bt$, in line with the studies of Refs.~\cite{Collins:2017oxh,Vladimirov:2020umg} based on more general formal considerations.
 
 The large $\bt$ behaviour of  our models, supplemented with  constraints from BELLE data, seems to be well supported by the lattice determinations of the CS kernel from quasi TMD wave functions~\cite{Schlemmer:2021aij, Li:2021wvl,Shanahan:2021tst,LPC:2022ibr}, which evidence the slow variation  of the kernel in this region of $\bt$. Remarkably, our extractions are in very good agreement with the calculations of  Refs.~\cite{Shanahan:2021tst,LPC:2022ibr} where an NLO matching is applied.  This is a very important cross-check, as lattice QCD calculations are based on totally different and independent  methodologies.
 
 On the other hand, little can be inferred from this analysis about the small-$\bt$ behaviour of the CS kernel and of $\gk$. This might be at least partially due to the relatively low energy of the BELLE experiment, but this is an issue which deserves more extensive studies,  including higher accuracy in the perturbative expansion. A more rigorous formal treatment will be presented in  Ref.~\cite{Boglione-Simonelli:2022}.

 A very important theoretical consideration regards the transition between short and long distance behaviour, which should be carefully treated when embedding models into the type of TMD FF definition like that of \eref{tmd_NLL}, where the small $\bt$ behaviour is, in principle, constrained by collinear factorization. In general, such constraints are not guaranteed unless models are optimally embedded, especially at small and moderate scales. Recently, this and related issues have been comprehensively addressed in Ref.~\cite{Gonzalez-Hernandez:2022ifv} where, based on theoretical considerations, a practical recipe for phenomenology was provided that allows a more reliable combination of models of nonperturbative behaviour into the CSS formalism.  These considerations will likely help to resolve some of the issues we found at small $\bt$ in our analysis. We plan to pursue this techniques in future work.

Another relevant aspect concerns the estimation of the errors affecting the phenomenological extraction of TMDs from experimental data.
It is important to stress that while statistical errors do provide insight into the precision with which TMDs can be extracted, theoretical errors play also an important role, which remarkably affect accuracy. We addressed  two sources  of such  errors and provided rough estimates of their size. First, we  considered  the effect that the statistical errors of the  collinear functions have  in the extraction of the unpolarized TMD FF, by refitting our model with each one of the sets provided by the NNFF collaboration. Second, the use of two different models for $\gk$ allowed us to assess how profile functions extracted depend on model choices, as seen in \fref{mdktilde}. In both cases, our estimates are meant to provide examples  of how important it is to perform error estimation beyond statistical uncertainties. More work is needed in order to address these issues with a more robust approach.

A possible future improvement in our analysis regards the treatment of experimental errors. For our work, we added in quadrature all errors provided by the BELLE collaboration which may be a matter of concern, specially regarding correlated systematic errors, since they should be treated on a different footing. This can be done, for instance, by introducing nuisance parameters in  the $\chi^2$ statistic, in the form of a shift to  theoretical estimates. This, however, likely requires more detailed information about the different sources of correlated systematic uncertainties. In our case, attempting to employ such methodology resulted in large values of the minimal $\chi^2$ although rendering almost identical results in the profile functions. 

Although our analysis was carried out on a rather limited subset of the BELLE data, we consider this work an essential first step.
We stress that, to the best of our knowledge, this is the only phenomenological analysis where the thrust dependence of the cross section is explicitly taken into account and well described over three different bins. Other studies~\cite{Kang:2017glf,DAlesio:2020wjq} resort to a combination of the thrust bins, resulting in a cross section which is some sort of average over thrust, or simply integrate it over. 
Extending our results to a wider range of thrust and $\zh$ bins requires further formal developments on identifying and extending the optimal kinematic region where the TMD formalism developed for region 2 in $e^+e^-\to h X$ can be successfully applied\cite{Boglione:2021wov,Boglione:2021vug}. Moreover, the connection between the regularization of the rapidity divergences and the thrust dependence must be set on a more solid formal ground, as it crucially affects the correlation among $T$, $P_T$ and $\zh$.
This will likely improve the quality of the extraction by allowing to possibly include more data points while achieving an even better  agreement to data~\cite{Boglione-Simonelli:2022}.

\section*{Acknowledgements}

We thank Ted Rogers for useful discussions regarding formal aspects of the CS kernel.  
We are grateful to Alexey Vladimirov for making available to us the results of the SV19 extraction  and the SVZES lattice calculation of the CS kernel, 
and to Andrea Signori for providing results of the PV19 fit.
We thank Xu Feng, Michael Wagman and Qi-An Zhang for interesting discussions on their lattice QCD determination of the Collins-Soper kernel and for providing us with relevant recent results on the subject.\\ This project has received funding from the European Union’s Horizon 2020 
research and innovation programme under grant agreement No 824093.

\appendix

\section{Wilson coefficients and $g_i$ functions \label{app:tmd}}

In this Appendix we provide the explicit expression of the quantities necessary to compute the perturbative part of the TMD FF.    
 The 1-loop Wilson coefficients $\mathcal{C}^{[1]}(z) \equiv \mathcal{C}^{[1]}(z,b_*;\mu_{b_*},\mu_{b_*}^2)$ appearing in \eref{tmd_NLL} are calculable in  pQCD and are given by ~\cite{Collins:2011zzd}
    \begin{align}
    z^2 \, \mathcal{C}^{[1]}_{q/q}(z) &=
    2 C_F \, \left(
    1-z + 2 \, \frac{1+z^2}{1-z} \,
    \log{z}
    \right)\nonumber\\
    & \quad - C_F \frac{\pi^2}{6} \delta(1-z)
    \label{eq:Wqq_1loop}\\
    z^2 \, \mathcal{C}^{[1]}_{g/q}(z) &= 2 C_F  \, \left[
    z + 2 \, \frac{1+(1-z)^2}{z} \, \log{z}
    \right].
    \label{eq:Wgq_1loop}
    \end{align}
To reach NLL accuracy the anomalous dimension $\gamma_K$ of the soft kernel is expanded up to 2-loops, while all other quantities are written to 1-loop.
The functions $g_i$, $i=1,2$ and $g^K_j$, $j=2,3$, required to reach NLL-accuracy in the expression of the TMD FF of \eref{tmd_NLL}, depend on the variable $\lambda = 2\,\beta_0 \, 
a_S(Q) \, \log{\frac{Q}{\mu_{b_\star}}}$ and are given by~\cite{Boglione:2020auc}:
%
    %
    \label{eq:gi_sud}
    \begin{flalign}
    &g_1(\lambda) = 
    \frac{\gamma_K^{[1]}}{4\beta_0} \left( 
    1+\frac{\log{\left(1 - \lambda\right)}}{\lambda}
    \right);
    \label{eq:g1tmd}&& \\
    &g_2(\lambda) = 
    \frac{\gamma_K^{[1]}}{8\beta_0^2}\,
    \frac{\beta_1}{\beta_0}\,
    \frac{\lambda}{1 - \lambda}\,
    \left( 
    1+\frac{\log{\left(1-\lambda\right)}}{\lambda} \right.
    \notag &&\\
    & \hspace{3.9cm} \left. +\frac{1}{2}\,
    \frac{1 - \lambda}
    {\lambda}\,
    \log^2{\left(1 - \lambda\right)}
    \right) - 
    \notag &&\\
    &\quad-\frac{\gamma_K^{[2]}}{8\beta_0^2}\,
    \left( 
    \frac{\lambda}{1-\lambda}+\log{\left(1-\lambda\right)}
    \right) - 
    \frac{\gamma_d^{[1]}}{2\beta_0}\,\log{\left(1-\lambda\right)};
    \label{eq:g2tmd}&& \\
    &g_1^{\text{K}}(\lambda) = 
    \frac{\gamma_K^{[1]}}{2\beta_0}\log{\left(1-\lambda\right)};
    \label{eq:g1csker}&& \\
    &g_2^{\text{K}}(\lambda) = 
     \frac{\gamma_K^{[1]}}{4\beta_0^2}\,
     \frac{\beta_1}{\beta_0}\,
     \frac{\lambda^2}{1-\lambda}
     \left(1 + \frac{\log{\left(1-\lambda\right)}}{\lambda} \right) -
     \frac{\gamma_K^{[2]}}{4\beta_0^2}\,
     \frac{\lambda^2}{1-\lambda}\,,
     \label{eq:g2csker}&&
    \end{flalign}
    %
%
with
\begin{subequations}
\label{eq:gammaK_loops}
\begin{align}
&\gamma_K^{[1]} = 16\,C_F ;\\
&\gamma_K^{[2]} = 2 \, C_A \, C_F \,
\left(
\frac{536}{9} - \frac{8 \pi^2}{3}
\right) -
\frac{160}{9} \, C_F \, n_f\,,
\end{align}
\end{subequations}
where $n_f$ is the total number of fermion fields  considered while 
$\beta_0$ and $\beta_1$ are the coefficients of the beta functions up to $2$ loops:
\begin{subequations}
\label{eq:beta_2loop}
\begin{align}
&\beta_0 = \frac{11}{3}C_A - \frac{2}{3}n_f,\\
&\beta_1 =  \frac{34}{3}C_A^2 -
\frac{10}{3}C_A \, n_f - 2 C_F n_f. 
\end{align}
\end{subequations}
We refer to Ref.~\cite{Collins:2017oxh} for the explicit values of the anomalous dimensions, Collins-Soper kernel and the QCD beta function coefficients, having care of multiplying by $2$ all the coefficients related to the CS-kernel.


%

\end{document}